\documentclass[a4paper,10pt]{article}
\usepackage{amssymb,amsmath,amsfonts,placeins,pbox,multirow,mathtools,bbold}
\usepackage{graphicx,rotate,color,slashed,cite,epstopdf,verbatim,url,multirow,booktabs}
\usepackage{epstopdf}
\usepackage{subfig}
\usepackage{graphicx}
\usepackage{array}
\renewcommand{\arraystretch}{1.5} 
\usepackage[a4paper, left=2cm, right=2cm, top=2cm, bottom=2cm]{geometry}

\usepackage[colorlinks=true,
            linkcolor=magenta,
            urlcolor=blue,
            citecolor=blue]{hyperref}
\usepackage[utf8x]{inputenc}

\usepackage[capitalise]{cleveref}
\usepackage[dvipsnames]{xcolor}

\crefname{section}{Sec.}{Secs.}
\crefname{table}{Tab.}{Tabs.}
\crefname{figure}{Fig.}{Figs.}
\crefname{equation}{Eq.}{Eqs.}
\crefname{appendix}{Appendix}{Appendix}

\numberwithin{equation}{section}

\def\ssbh#1//#2//{\ensuremath{\xrightarrow [\substack{#2}]
    {\parbox{3cm}{\hfil $\scriptstyle \langle #1 \rangle$ \hfil}}}}

\let\CapTion=\caption
\def\caption#1{\CapTion{\em #1}}

\title{\bf Heavy vector-like quarks decaying to exotic scalars:\\ a case study with triplets}

\author{\bf Avik Banerjee$^{a}$\thanks{avik@chalmers.se}\,, Venugopal Ellajosyula$^{b}$\thanks{venugopal.ellajosyula@physics.uu.se}\, and Luca Panizzi$^{b,c,d,e}$\thanks{luca.panizzi@unical.it}
\\ \medskip 
{}$^{a}$\small\em Department of Physics, Chalmers University of Technology, Fysikg\aa rden, 41296 G\"oteborg, Sweden\\[-15pt]
{}$^{b}$\small\em Department of Physics and Astronomy, Uppsala University, Box 516, SE-751 20 Uppsala, Sweden\\[-8pt]
{}$^{c}$\small\em Dipartimento di Fisica, Universit\`a della Calabria, I-87036 Arcavacata di Rende, Cosenza, Italy\\[-8pt]
{}$^{d}$\small\em INFN-Cosenza, I-87036 Arcavacata di Rende, Cosenza, Italy.\\[-9pt]
${}^{e}$\small\em School of Physics and Astronomy, University of Southampton, Highfield, Southampton SO17 1BJ, UK.
}
\date{}

\begin{document}

\maketitle

% % % % % % % % % % % % % % % % Abstract % % % % % % % % % % % % % % % % % 

\begin{abstract}
We investigate the pair production of a vector-like quark triplet with hypercharge 5/3 decaying into top quark and a complex scalar triplet with hypercharge 1 at the LHC. This novel scenario, featuring particles with exotic charges - two quarks with charge 8/3 and 5/3 and a scalar with charge 2 - serves as a unique window to models based on the framework of partial compositeness, where these particles naturally emerge as bound states around the TeV scale. Leveraging on the LHC data we establish exclusion limits on the masses of the vector-like quark and the scalar triplet. Subsequently, we design an analysis strategy aimed at improving sensitivity in the region which is still allowed. Our analysis focuses on two specific regions in the parameter space: the first entails a large mass gap between the vector-like quarks and the scalars, so that the vector-like quarks can decay into the scalars; the second involves a small mass gap, such that this decay is forbidden. To simplify the parameter space, both vector-like quarks and scalars are assumed to be degenerate or almost degenerate within the triplets, such that chain decays between fermions and scalars are suppressed. As a result, we found that final states characterized by a same-sign lepton pair, multiple jets, and high net transverse momentum ({\it i.e.} effective mass) will play a pivotal role to unveil this model and, more in general, models characterised by multiple vector-like quarks around the same mass scale during the high luminosity LHC phase. 
\end{abstract}
% % % % % % % % % % % % % % % % % % % % % % % % % % % % % % % % % % % % % %

\bigskip

% % % % % % % % % % % % % % % Main Document% % % % % % % % % % % % % % % % 
\newpage
\tableofcontents
\bigskip
\hrule
\bigskip
% % % % % % % % % % % % % % % % % % % % % % % % % % % % % % % % % % % % % %
\section{Introduction}
\label{intro}
% % % % % % % % % % % % % % % % % % % % % % % % % % % % % % % % % % % % % %

At present the Large Hadron Collider (LHC) is running in its third operational phase, dedicated to search for new physics beyond the Standard Model (BSM) by colliding proton-proton beams at a center-of-mass energy of 13.6 TeV. From the theory side, the electroweak hierarchy problem has been a significant guiding factor to motivate BSM physics for quite some time. Given the close connection between the hierarchy problem, the Higgs boson, and the top quark, majority of BSM theories aimed at resolving this issue involve an expanded Higgs and/or top quark sectors. In other words, these BSM scenarios often predict the existence of extra spin-1/2 and spin-0 particles with masses around or above the TeV scale. However, no significant excess over the Standard Model (SM) background has been reported so far by the ATLAS or CMS collaborations, thereby considerably narrowing the room for simple BSM extensions. 

Our primary motivation for this work stems from partial compositeness models with a pseudo-Nambu Goldstone (pNGB) Higgs boson, which were proposed to address the hierarchy problem and elucidate the origin of the top quark mass \cite{Kaplan:1983fs, Kaplan:1991dc}. A key prediction of these models is the existence of vector-like quarks (VLQs) and new composite pNGBs with electroweak quantum numbers. Vector-like quarks are color triplet fermions whose left- and right- handed chiralities transform identically under the SM gauge group. VLQs also appear as the higher Kaluza Klein (KK) modes in the holographic realizations of composite Higgs models \cite{Antoniadis:1990ew,Contino:2003ve,Csaki:2003sh,Agashe:2004rs,Erdmenger:2020lvq,Erdmenger:2020flu}. Other motivated extensions of the SM, such as the two Higgs doublet models \cite{Benbrik:2019zdp,Bonilla:2021ize} or models with extended gauge symmetries  \cite{ArkaniHamed:2002qx,Abbas:2017vle} may also feature new scalars and vector-like quarks to tackle challenges in flavor physics, providing additional sources of CP violation and so forth \cite{Branco:2021vhs,Botella:2021uxz}. 

VLQs have been widely studied phenomenologically in the context of simplified models where they only interact with SM particles\cite{Lavoura:1992np,delAguila:2000rc,AguilarSaavedra:2005pv,Cynolter:2008ea,AguilarSaavedra:2009es,Mrazek:2009yu,Cacciapaglia:2012dd,Berger:2012ec,Okada:2012gy,DeSimone:2012fs,Vignaroli:2012nf,Falkowski:2013jya,Buchkremer:2013bha,AguilarSaavedra:2013qpa,Matsedonskyi:2014mna,Matsedonskyi:2015dns,Barducci:2017xtw,Panella:2017spx,Buckley:2020wzk,Deandrea:2021vje,Belyaev:2021zgq,Arsenault:2022xty}. Pair production of VLQs at hadron collider is mainly governed by QCD interactions, leading to a cross-section which only depends on the mass of the VLQs and the centre-of-mass energy of the collider. Extensive searches for VLQs have been conducted at the LHC by both ATLAS and CMS collaborations, resulting in stringent bounds on the VLQ mass surpassing the TeV scale, irrespective of their specific decays into SM particles 
\cite{Sirunyan:2018yun,ATLAS:2018uky,Aaboud:2018xpj,Aaboud:2018uek,Aaboud:2018saj,Aaboud:2018xuw,ATLAS:2022hnn,ATLAS:2022ozf,ATLAS:2022tla,Sirunyan:2018qau,Sirunyan:2017pks,Sirunyan:2018omb,Sirunyan:2019sza,Sirunyan:2020qvb,CMS:2021iuw,CMS:2022yxp,CMS:2022fck,CMS:2023agg}.
However, two aspects have to be taken into account when interpreting current bounds:
\begin{enumerate}
\item the VLQs typically also interact with light BSM particles present in partial compositeness and other motivated new physics models \cite{Benbrik:2019zdp,Banerjee:2016wls,Chala:2018qdf,Bizot:2018tds,Xie:2019gya,Ramos:2019qqa,Aguilar-Saavedra:2019ghg,Cacciapaglia:2019zmj,Dasgupta:2019yjm,Wang:2020ips,Dermisek:2020gbr,Dermisek:2021zjd,Dasgupta:2021fzw,Corcella:2021mdl,Bhardwaj:2022nko,Bhardwaj:2022wfz,Verma:2022nyd,Belyaev:2022shr};
\item in such theory-motivated scenarios, there are usually more than one VLQ, with different charges and often in a similar mass range, especially when belonging to the same multiplet.
\end{enumerate}
These elements can, on the one hand, significantly alter the bounds obtained so far under the assumption that the SM is extended with one VLQ multiplet, and its components can only decay to SM particles. On the other hand, potentially observable signal events may arise from the production and decay of more than one VLQ, especially if the VLQs have high masses, such that their individual observation at the LHC is not possible even in the high-luminosity phase of the LHC (HL-LHC).\\   

In this paper, we explore a novel phenomenological scenario where the SM is extended by two $SU(2)_L$ triplets: a VLQ triplet with hypercharge $Y=5/3$ ($Q\in{\bf 3}_{5/3}$), and a complex scalar triplet with $Y=1$ ($S\in {\bf 3}_{\pm 1}$), respectively. The VLQ ${\bf 3}_{5/3}$ is composed of new colored fermionic states with electric charge $2/3$, $5/3$ and $8/3$, while the colorless scalar triplet contains an electrically neutral, a charged, and a doubly charged particle. Among the composite Higgs models, two minimal examples accommodating a VLQ  triplet with hypercharge $Y=5/3$ are given by the coset structures $SO(5)/SO(4)$ \cite{Contino:2003ve,Agashe:2004rs,Contino:2006qr,Panico:2011pw,Contino:2011np,Azatov:2011qy,DeSimone:2012fs,Matsedonskyi:2012ym,Marzocca:2012zn,Pomarol:2012qf,Panico:2012uw,Carena:2014ria,Montull:2013mla,Carmona:2014iwa,Panico:2015jxa,Niehoff:2015iaa,Kanemura:2016tan,Gavela:2016vte,Banerjee:2017wmg,Liu:2017dsz} and $SU(5)/SO(5)$ \cite{Golterman:2015zwa,Ferretti:2014qta,Ferretti:2016upr,Agugliaro:2018vsu,Banerjee:2022izw,Cacciapaglia:2021uqh}. In both the cases, ${\bf 3}_{5/3}$ arises from the symmetric irrep of the unbroken global symmetry (see \cref{model_comparison}). The coset $SO(5)/SO(4)$, also known as the minimal composite Higgs model (MCHM) leads to a pNGB Higgs doublet alone, while $SU(5)/SO(5)$ coset has a richer pNGB spectra including a complex scalar triplet. The phenomenology of a VLQ ${\bf 3}_{5/3}$ arising from the MCHM and decaying exclusively into SM final states has been discussed in \cite{Matsedonskyi:2014lla}.  

\begin{table}[t!]
%\begin{small}
\def\arraystretch{1.0}
	\begin{center}
\begin{tabular}{ccccccccc}
\toprule\toprule
\parbox{2.5cm}{Coset ($G/H$)} & \parbox{5.5cm }{\centering VLQ (irrep under $H$)} & \parbox{4.5cm }{\centering pNGB (irrep under $H$)} \\
\midrule
$\rm\frac{SO(5)}{SO(4)}\times U(1)_X$ & \parbox{3.5cm}{ \begin{align*} 9_{2/3} &\to (3,3)_{2/3} \to 3_{-1/3}+3_{2/3}+\textcolor{red}{3_{5/3}} \end{align*}} & \parbox{4cm}{\begin{align*} 4 &\to (2,2)\to 2_{\pm 1/2} \end{align*}} \\
\midrule
$\rm\frac{SU(5)}{SO(5)}\times U(1)_X$ & \parbox{3.5cm}{ \begin{align*} 14_{2/3} &\to (1,1)_{2/3}+(2,2)_{2/3}+(3,3)_{2/3} \\ &\to 1_{2/3}+2_{1/6}+2_{7/6} + 3_{-1/3}+3_{2/3}+\textcolor{red}{3_{5/ 3}} \end{align*}} & \parbox{3.cm}{\begin{align*} 14 &\to (1,1)+(2,2)+(3,3) \\ &\to 1_{0}+2_{\pm 1/2}+3_{0}+\textcolor{red}{3_{\pm 1}}\end{align*}} \\
\bottomrule\bottomrule
\end{tabular}
\caption{\small\it VLQ and pNGB contents of the SO(5)/SO(4) and SU(5)/SO(5) cosets. The $H$ irreps are decomposed under $H\times$U(1)$_X\to$SU(2)$_L\times$SU(2)$_R\times$U(1)$_X\to$SU(2)$_L\times$U(1)$_Y$, where $Y=T^3_R+X$. Among the pNGBs, the $(2,2)$ denotes the usual Higgs doublet.}
\label{model_comparison}
\end{center}
%\end{small}
\end{table}

We consider non-standard decays of the VLQ ${\bf 3}_{5/3}$ into a top quark and a complex scalar triplet, which is plausible in the $SU(5)/SO(5)$ coset.\footnote{We mention in passing that the $SU(5)/SO(5)$ coset belongs to the class of models that can originate from a 4D confining gauge theory with only fermionic matter in the ultra-violet \cite{Ferretti:2013kya,Ferretti:2014qta,Ferretti:2016upr}.} The production and decay of the VLQ ${\bf 3}_{5/3}$ are dominated by its interaction with the SM particles and the complex scalar triplet, while its interactions with other BSM multiplets, shown in \cref{model_comparison}, do not have appreciable impact. Thus for the sake of simplicity, we perform the analysis by extending the SM with a VLQ ${\bf 3}_{5/3}$ and a scalar ${\bf 3}_{\pm 1}$ only, rather than considering the full particle content of the $SU(5)/SO(5)$ coset. We construct a simplified Lagrangian to characterize the pertinent interactions, allowing the coupling strengths to vary as free parameters. Subsequently, we leverage insights from the partial compositeness framework to provide an order-of-magnitude estimate for these coupling strengths. 

Our analysis, utilizing a simplified Lagrangian, possesses an intriguing characteristic. We can draw certain inferences, {\em albeit} approximate, for both  $SU(5)/SO(5)$  and  $SO(5)/SO(4)$ cosets, depending on whether the mass gap between the VLQ and the scalar triplet is greater or smaller than the top quark mass. In the former case, VLQs predominantly undergo 2-body decays into the scalar triplet and top quark, closely resembling the situation in the $SU(5)/SO(5)$ composite Higgs model. Conversely, in the case of a small mass gap, VLQs decay primarily into SM final states, aligning more with the MCHM scenario. 

Exclusion limits on the VLQ and the scalar masses are obtained by recasting publicly accessible Run 2 data from a set of ATLAS and CMS searches at the LHC. It is worth highlighting that the existence of a VLQ with an electric charge of $8/3$ (referred to as $Y_{8/3}$) and its primary decay mode into a top quark and a doubly charged scalar ($t+S^{++}$) introduces entirely novel search topologies at the LHC. We study the prospect of using such new topologies characterized by same sign lepton (SSL) pairs, and abundant jet multiplicities in the final state, to search for the triplet VLQs decaying into BSM scalars at HL-LHC. 

The paper is organized as follows: in \cref{theory} we describe the model, provide an estimate of the coupling strengths inspired by the partial compositeness scenarios, and discuss the branching ratio patterns of the VLQs and BSM scalars. The constraints on the VLQ and scalar masses are obtained in \cref{sec:LHCconstraints} by recasting limits from existing experimental searches at the LHC. The discovery prospect for the VLQs decaying to the exotic scalars at the HL-LHC is discussed in \cref{sec:HL_LHC_prospect}. Finally, we conclude in \cref{conc}.

% % % % % % % % % % % % % % % % % % % % % % % % % % % % % % % % % % % % % %
\section{Model description}
\label{theory}
% % % % % % % % % % % % % % % % % % % % % % % % % % % % % % % % % % % % % % 

We extend the SM Lagrangian by adding new gauge invariant terms with a vector-like quark triplet ${\bf 3}_{5/3}$ under the  SU(2)$_L\times$U(1)$_Y$, denoted by $Q\equiv (Y_{8/3},X_{5/3},T_{2/3})$, and a complex scalar triplet ${\bf 3}_{\pm 1}$, denoted by $S\equiv (S^{\pm\pm},S^\pm,S^0)$. 
The relevant pieces of the new physics Lagrangian ($\mathcal{L}_{\rm NP}$) are given by
\begin{align}
\mathcal{L}_{\rm  NP} &= \mathcal{L}_{ Q^2+S^2} + \mathcal{L}_{Q} + \mathcal{L}_{S}\,,
\label{L_NP}
\end{align}
where
\begin{align}
\label{L_quad}
\mathcal{L}_{Q^2+S^2} &= \bar{Q}\left(i\slashed D - m_Q\right) Q
+ \left(|D_\mu S|^2 -m_{S}^2 |S|^2\right),\\
\nonumber\\
\mathcal{L}_{Q} &= \frac{e}{\sqrt{2}s_W}\left[
\kappa^W_{T,L}\bar{T}_{2/3}\slashed W^+ P_L b+ \kappa^W_{X,L}\bar{X}_{5/3}\slashed W^+ P_L t + L\leftrightarrow R \right] + {\mathrm{h.c.}}\nonumber\\
& + \frac{e}{s_W c_W}\left[\kappa^Z_{T,L}\bar{T}_{2/3}\slashed Z P_L t + L\leftrightarrow R \right] + {\mathrm{h.c.}} + h\left[ \kappa^{h}_{T,L} \bar{T}_{2/3} P_L t + L\leftrightarrow R\right] + {\mathrm{h.c.}}  \label{L_Q}\\
\nonumber\\
\mathcal{L}_{S} &= S^0\left[\lambda^{S^0}_{t,L} \bar{t} P_L t + \lambda^{S^0}_{b,L} \bar{b} P_L b + \kappa^{S^0}_{T,L} \bar{T}_{2/3} P_L t + \kappa^{S^0}_{TT,L} \bar{T}_{2/3} P_L T_{2/3} + L\leftrightarrow R\right] + {\mathrm{h.c.}} \nonumber\\
\nonumber
&+ S^{++}\left[\kappa^{ S^{++}}_{Y,L} \bar{Y}_{8/3} P_L t + \kappa^{ S^{++}}_{X,L} \bar{X}_{5/3} P_L b + \kappa^{ S^{++}}_{YT,L} \bar{Y}_{8/3} P_L T_{2/3} + L\leftrightarrow R\right]  + {\mathrm{h.c.}} \nonumber\\
&+ S^+\left[\lambda^{S^+}_{L} \bar{t} P_L b + \kappa^{S^+}_{X,L} \bar{X}_{5/3} P_L t + \kappa^{S^+}_{T,L} \bar{T}_{2/3} P_L b + \kappa^{S^+}_{XT,L} \bar{X}_{5/3} P_L T_{2/3} \right.  \nonumber\\ 
& \left. + \kappa^{S^+}_{YX,L} \bar{Y}_{8/3} P_L X_{5/3} + L\leftrightarrow R\right] + {\mathrm{h.c.}} \label{L_s}
\end{align}
The covariant derivatives in \eqref{L_quad} involve QCD (for the VLQs only) and electroweak gauge interactions. The coefficients $\lambda$, $\kappa$ and the masses $m_Q$, $m_S$ are arbitrary free parameters of the model, while $s_W\equiv\sin\theta_W$ denotes the Weinberg angle. We assume that only the Higgs doublet receives a vacuum expectation value (VEV) $v$, ensuring tree-level custodial invariance. Consequently, the 125 GeV Higgs boson ($h$) does not mix with $S^0$, the neutral component of the scalar triplet. In addition, motivated by the composite Higgs models, we assume that the VLQs and the triplet scalar couples only to the third generation quarks. 

Taking cue from the partial compositeness scenarios, in \cref{coups} we present an order of magnitude estimate for the coupling strengths $\lambda$ and $\kappa$ in powers of $v/f$, where $f\sim 1$ TeV denotes the decay constant of the pNGB Higgs boson in such models. 
\begin{table}[h!]
%\begin{small}
\def\arraystretch{1.2}
	\begin{center}
\begin{tabular}{cccccccccccccccccccccc}
\toprule\toprule
$\lambda^{S^+}_L$ & $\lambda^{S^+}_R$ & $\lambda^{S^0}_{t,L}$ & $\lambda^{S^0}_{t,R}$ & $\lambda^{S^0}_{b,L}$ & $\lambda^{S^0}_{b,R}$ & $\kappa^{S^+}_{X,L}$ & $\kappa^{S^+}_{X,R}$ \\
\midrule
$\mathcal{O}(v/f)$ & $\mathcal{O}(v/f)$ & $\mathcal{O}(v/f)$ & $0$ & $0$ & $\mathcal{O}(v/f)$ & $\mathcal{O}(v/f)$ & $\mathcal{O}(1)$ \\
\bottomrule\toprule
$\kappa^{S^+}_{T,L}$ & $\kappa^{S^+}_{T,R}$ & $\kappa^{S^{++}}_{Y,L}$ & $\kappa^{S^{++}}_{Y,R}$ & $\kappa^{S^{++}}_{X,L}$  & $\kappa^{S^{++}}_{X,R}$ & $\kappa^{S^0}_{T,L}$ & $\kappa^{S^0}_{T,R}$ \\ 
\midrule
$\mathcal{O}(v/f)$ & $0$ & $0$ & $\mathcal{O}(1)$ & $\mathcal{O}(v/f)$ & $0$ & $\mathcal{O}(v/f)$ & $\mathcal{O}(1)$ \\
\bottomrule\toprule
$\kappa^{h}_{T,L}$ & $\kappa^{h}_{T,R}$ & $\kappa^{W}_{T,L}$ & $\kappa^{W}_{T,R}$ & $\kappa^{W}_{X,L}$ & $\kappa^{W}_{X,R}$ & $\kappa^{Z}_{T,L}$  & $\kappa^{Z}_{T,R}$ \\ 
\midrule
$\mathcal{O}(v^2/f^2)$ & $\mathcal{O}(v/f)$ & $0$ & $0$ & $0$ & $\mathcal{O}(v^2/f^2)$ & $0$ & $\mathcal{O}(v^2/f^2)$ \\
\bottomrule\toprule
$\kappa^{S^+}_{XT,L}$ & $\kappa^{S^+}_{XT,R}$ & $\kappa^{S^+}_{YX,L}$ & $\kappa^{S^+}_{YX,R}$ & $\kappa^{S^{++}}_{YT,L}$ & $\kappa^{S^{++}}_{YT,R}$ & $\kappa^{S^0}_{TT,L}$ & $\kappa^{S^0}_{TT,R}$ \\ 
\midrule
$0$ & $\mathcal{O}(v^2/f^2)$ & $0$ & $0$ & $0$ & $\mathcal{O}(v^2/f^2)$ & $0$ & $\mathcal{O}(v^2/f^2)$ \\
\bottomrule\bottomrule
\end{tabular}
\caption{\small\it Partial compositeness inspired estimates for $\lambda$ and $\kappa$ appearing in the Lagrangian (\ref{L_Q}) and (\ref{L_s}) up to $\mathcal{O}(v^2/f^2)$.}
\label{coups}
\end{center}
%\end{small}
\end{table}
Notably, due to the triplet nature, the VLQs couple to the new scalars and a right-handed top quark with order one coupling strength, while the couplings of the VLQs with a $t$ or $b$ quark and the $W^\pm,Z$ bosons or the Higgs boson are suppressed with powers of $v/f$.  In the composite models, the pNGB scalars can also have higher dimensional Wess-Zumino-Witten anomaly interactions, which arise at one loop \cite{Ferretti:2016upr}. However, we neglect these interactions since they have negligible impact on our analysis. In \cref{EFT}, we provide an effective field theory (EFT) construction that leads to the phenomenological Lagrangian present in \cref{L_NP,L_quad,L_Q,L_s}, and justifies our choice of benchmark parameters for the collider analysis.

A few comments are necessary regarding the mass spectra of the model. The triplet nature of the VLQ suggests that masses of its components are almost degenerate at tree-level (denoted by $m_Q$). Since the mixing between the top quark and the $T_{2/3}$ component of the triplet $Q$ is suppressed at $\mathcal{O}(v^2/f^2)$, we neglect the small tree-level corrections ($\Delta m/m_Q\sim \mathcal{O}(v^4/f^4)$) to the mass of $T_{2/3}$. We also consider that the scalars $S^0$, $S^+$ and $S^{++}$ are nearly degenerate and their common mass, denoted by $m_S$ is smaller than $m_Q$. As a result, the leading decay channels of the scalars involve only SM final state particles. 
A full calculation of the one-loop corrections to $m_Q$ and $m_S$ depends on the values of the model parameters, and the ultra-violate cut-off ($\Lambda\sim 4\pi f$) of the theory, and is beyond the scope of this paper.
In \cref{app:nondegenerate} we will however discuss the impact of lifting the degeneracy assumption on the numerical results by considering also a scenario where the masses of VLQs and scalars within the triplets are artificially split.

% % % % % % % % % % % % % % % % % % % % % % % % % % % % % % % % % % % % % %
\subsection{Branching ratio patterns}
\begin{figure}[h!]
\begin{center}
            \includegraphics[width=\linewidth]{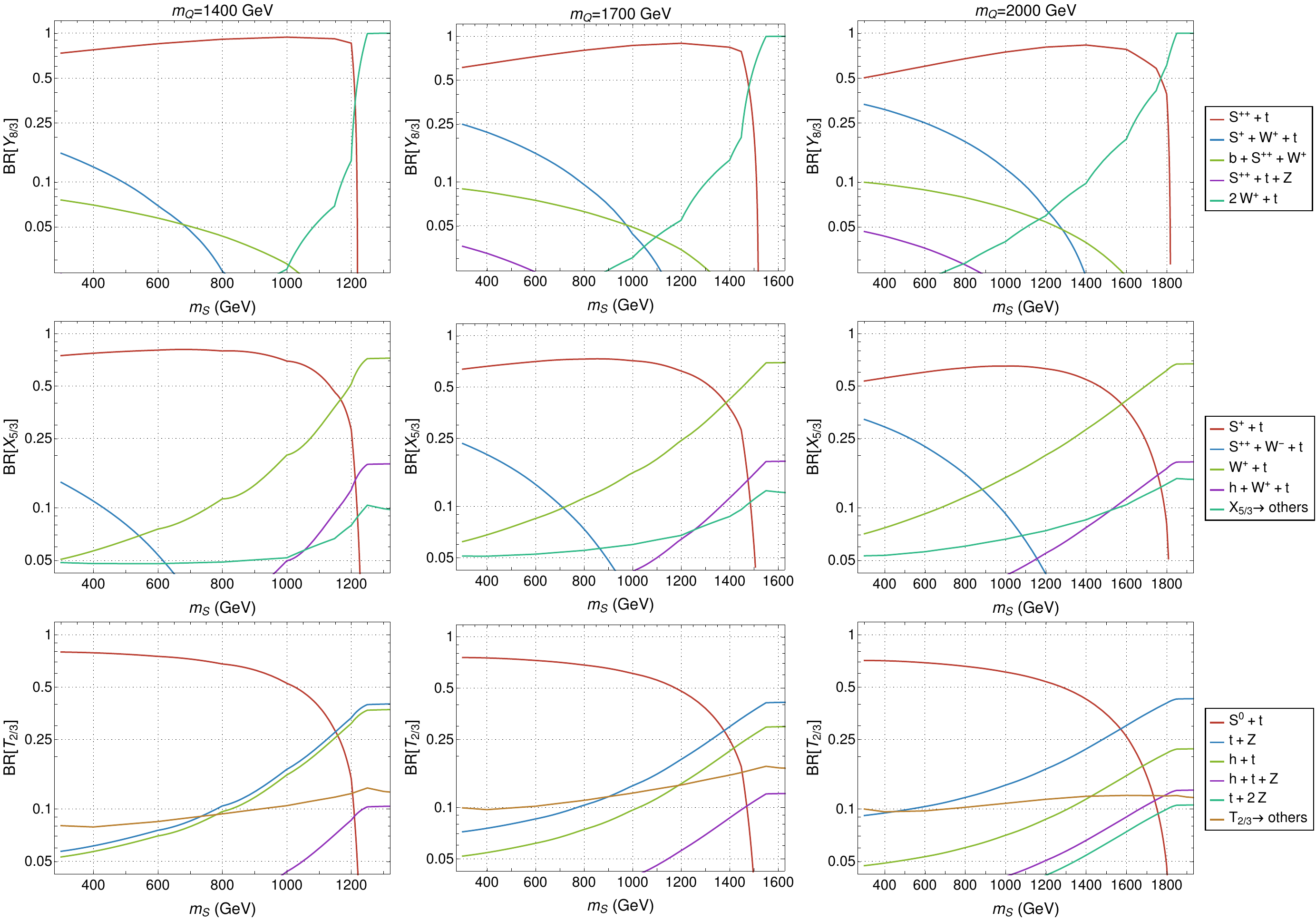} %\hspace{10pt}
\end{center}    	
    \caption{\sf\it BRs of $Y_{8/3}$ (top), $X_{5/3}$ (middle)  and $T_{2/3}$ (bottom) into leading 2-body and 3-body decay channels as a function of $m_S$, for $m_Q=1400$ (left), $1700$ (middle), and $2000$ (right) GeV. $X_{5/3}\to {\rm others}$ and $T_{2/3}\to {\rm others}$ denote all decay channels with $BR<0.1$.}
    \label{fig:vlq_br}
\end{figure}

% % % % % % % % % % % % % % % % % % % % % % % % % % % % % % % % % % % % % %
In \cref{decays} we present a list of the leading 2-body and 3-body decay channels of the VLQs, and the corresponding branching ratios (BRs) for $m_Q=1400,1700,$ and $2000$ GeV  as a function of $m_S$ are displayed in \cref{fig:vlq_br}. This mass range corresponds to the region where current bounds and HL-LHC projections are relevant, as will be described in \cref{sec:LHCconstraints}.
\begin{table}[h!]
%\begin{small}
\def\arraystretch{1.2}
	\begin{center}
\begin{tabular}{cccccccccccccccccccccc}
\toprule\toprule
VLQ & 2-body & & 3-body\\
\midrule
$Y_{8/3}$ & $(t \, + \, S^{++})$ & & $(t \, + \, W^{+} \, + \, W^{+} / S^{+})$, $(b \, + \,  S^{++} \, + \, W^{+} / S^{+})$ \\

&  & & $(t \, + \, S^{++} \, + \, Z)$ \\\\

\multirow{2}{*}{$X_{5/3}$} &  $(t \, + \, W^{+} / S^{+})$ & & $(t \, + \, S^{++} \, + \, W^{-} / S^{-})$, $(t \, + \, W^{+} \, + \, h / Z / S^{0})$ \\

& $(b \, + \, S^{++})$ & & $(b \, + \, S^{+} \, + \, W^{+} / S^{+})$, $(t \, + \, S^{+} \, + \, S^{0*})$, $(t \, + \, t \, + \, \bar{b})$ \\\\

\multirow{2}{*}{$T_{2/3}$} & $(t \, + \, h /Z / S^{0})$ & & $(t \, + h /Z / S^{0} \, + \, h /Z / S^{0})$, $(t \, + \, W^{-} \, + \, W^{+} / S^{+})$\\
 
 & $(b \, + \, S^{+})$ & & $(b \, + \, W^{-} \, + \, S^{++})$, $(t \, + \, b \, + \, \bar{b})$, $(t \, + \, t \, + \, \bar{t})$ & \\
\bottomrule\bottomrule
\end{tabular}
\caption{\small\it Leading 2-body and 3-body decay channels of the VLQs.
}
\label{decays}
\end{center}
%\end{small}
\end{table}

In \cref{coups_BP} we provide the benchmark values of the coupling strengths used to calculate the branching ratios and for the rest of the analysis, in consonance with the expectations from partial compositeness scenario. In \cref{EFT} we further explain how to obtain these benchmark values from the EFT construction.
\begin{table}[h!]
%\begin{small}
\def\arraystretch{1.2}
	\begin{center}
\begin{tabular}{cccccccccccccccccccccc}
\toprule\toprule
$\lambda^{S^+}_L$ & $\lambda^{S^+}_R$ & $\lambda^{S^0}_{t,L}$ & $\lambda^{S^0}_{t,R}$ & $\lambda^{S^0}_{b,L}$ & $\lambda^{S^0}_{b,R}$ & $\kappa^{S^+}_{X,L}$ & $\kappa^{S^+}_{X,R}$ \\
\midrule
$-0.123$ & $0.123$ & $0.174$ & $0$ & $0$ & $0.174$ & $-0.087$ & $1$ \\
\bottomrule\toprule
$\kappa^{S^+}_{T,L}$ & $\kappa^{S^+}_{T,R}$ & $\kappa^{S^{++}}_{Y,L}$ & $\kappa^{S^{++}}_{Y,R}$ & $\kappa^{S^{++}}_{X,L}$  & $\kappa^{S^{++}}_{X,R}$ & $\kappa^{S^0}_{T,L}$ & $\kappa^{S^0}_{T,R}$ \\ 
\midrule
$0.123$ & $0$ & $0$ & $1$ & $0.123$ & $0$ & $-0.174$ & $1$ \\
\bottomrule\toprule
$\kappa^{h}_{T,L}$ & $\kappa^{h}_{T,R}$ & $\kappa^{W}_{T,L}$ & $\kappa^{W}_{T,R}$ & $\kappa^{W}_{X,L}$ & $\kappa^{W}_{X,R}$ & $\kappa^{Z}_{T,L}$  & $\kappa^{Z}_{T,R}$ \\ 
\midrule
$0.015$ & $0.246$ & $0$ & $0$ & $0$ & $0.031$ & $0$ & $-0.043$\\
\bottomrule\toprule
$\kappa^{S^+}_{XT,L}$ & $\kappa^{S^+}_{XT,R}$ & $\kappa^{S^+}_{YX,L}$ & $\kappa^{S^+}_{YX,R}$ & $\kappa^{S^{++}}_{YT,L}$ & $\kappa^{S^{++}}_{YT,R}$ & $\kappa^{S^0}_{TT,L}$ & $\kappa^{S^0}_{TT,R}$ \\ 
\midrule
$0$ & $0.022$ & $0$ & $0$ & $0$ & $-0.022$ & $0$ & $-0.022$\\
\bottomrule\bottomrule
\end{tabular}
\caption{\small\it Benchmark couplings used in this paper, in accordance with the estimate given in \cref{coups}.}
\label{coups_BP}
\end{center}
%\end{small}
\end{table}

To discuss the BR patterns of the VLQs, we divide the $m_Q$ vs. $m_S$ plane in two regions:
$${\setlength{\arraycolsep}{2pt}\left\{\begin{array}{lll} \text{RL:} & m_Q-m_S > m_t & \text{(large mass gap)} \\ \text{RS:} & m_Q-m_S < m_t & \text{(small mass gap)} \end{array}\right.}\;.$$
In RL, $Y_{8/3}$ dominantly decays into the only possible 2-body final state $t+S^{++}$. The 3-body decay $Y_{8/3} \to t + W^+ + S^+$ gains in the BR at low $m_S$ with increasing values of $m_Q$. In RS, $Y_{8/3}$ decays exclusively into $t+W^+ + W^+$ via off-shell $X_{5/3}$ exchange \cite{Matsedonskyi:2014lla}. 

The $X_{5/3}$ decays into $t+S^+$ with maximum BR in RL, owing to the large coupling $\kappa^{S^+}_{X,R}\sim \mathcal{O}(1)$, followed by the 3-body decay channel $t+W^-+S^{++}$ for which the BR increases with $m_Q$. $X_{5/3} \to t+W^+$ is the leading decay channel in RS. In case of $T_{2/3}$, the largest BR is into the decay $T_{2/3}\to t+S^0$ in RL, since $\kappa^{S^0}_{T,R}\sim \mathcal{O}(1)$. As we move towards RS, the decay of $T_{2/3}$ into SM 2-body final states $t+h$ and $t+Z$ start dominating. 

Note that, despite the partial decay width of $X_{5/3}\to b + S^{++}$ is governed by comparable couplings with respect to $X_{5/3}\to t + W^{+}$, the exotic decay is sub-leading in RS due to strong phase space suppression. In RL the same decay is still subleading because the couplings of $\bar{X}_{5/3}tS^+$ is dominant. A similar argument also holds for $T_{2/3} \to b + S^+$ decay. It is also worthwhile to mention that the interaction strength of $\bar{T}_{2/3}bW^+$ vertex arises at $\mathcal{O}(v^3/f^3)$, suppressing $T_{2/3}\to b + W^+$ decay compared to other 2-body decay channels of $T_{2/3}$, in stark contrast with the cases having a $SU(2)_L$ singlet or doublet VLQ. Therefore, we only account for coupling strengths up to $\mathcal{O}(v^2/f^2)$.

\begin{figure}[h!]
\begin{center}
            \includegraphics[width=\linewidth]{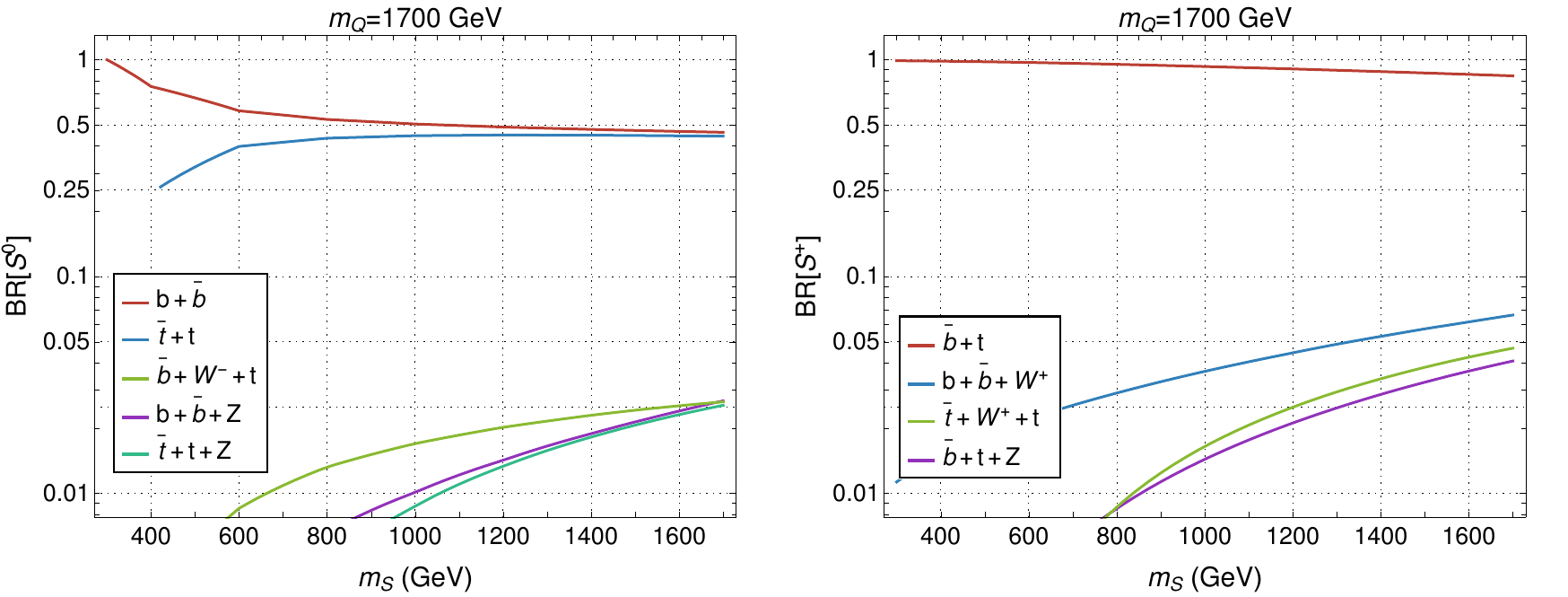} %\hspace{10pt}
\end{center}    	
    \caption{\sf\it BRs of $S^0$ (left), and $S^+$ (right) into SM final state particles as functions of $m_S$, for $m_Q=1700$ GeV. The $S^{++}$ always undergoes 3-body decay into $ W^+ t \Bar{b}$ with 100\% BR. The BRs of the scalars are approximately independent of the VLQ mass.}
    \label{fig:pngb_br}
\end{figure}

In \cref{fig:pngb_br}, the BRs of the scalars into different SM final states are shown as a function of $m_S$, keeping $m_Q=1700$ GeV. 
For the mass spectra considered in this paper, the main decay channels for the scalars accounting for more than 90\% of the branching ratios are
\begin{align}
        S^{++} \to W^+ t \bar{b}\,, %\, f_1 \bar{f_2} t \bar{b}\,, 
        \quad
        S^{+} \to t \bar{b}\,, \quad 
        S^0 \to b \bar{b}\,, \, t \bar{t}\,.
\end{align}
The partial decay widths of the scalars into SM final states are approximately independent of the VLQ mass. The electrically neutral scalar $S^0$ decays primarily into $t\Bar{t}$ and $b\Bar{b}$ final states with almost equal BRs at $m_S\gg m_t$, since $\lambda^{S^0}_{t,L}=\lambda^{S^0}_{b,R}$ and $\lambda^{S^0}_{t,R}=\lambda^{S^0}_{b,L}=0$. This case is different from the results presented in \cite{Banerjee:2022xmu,Cacciapaglia:2022bax} where the couplings of neutral pNGB scalar with the SM quarks are assumed to be proportional to the corresponding quark mass. Since the neutral scalar of the triplet we are considering does not acquire a VEV, here we consider more generic couplings, also justified from an EFT perspective shown in \cref{EFT}. The doubly charged scalar $S^{++}$ decays exclusively into the 3-body final state $W^+t\Bar{b}$ via an off-shell $S^+$ exchange, while the leading decay channel of $S^{+}$ is $S^+\to t \Bar{b}$. 

\begin{figure}[h!]
\begin{center}
            \includegraphics[width=\linewidth]{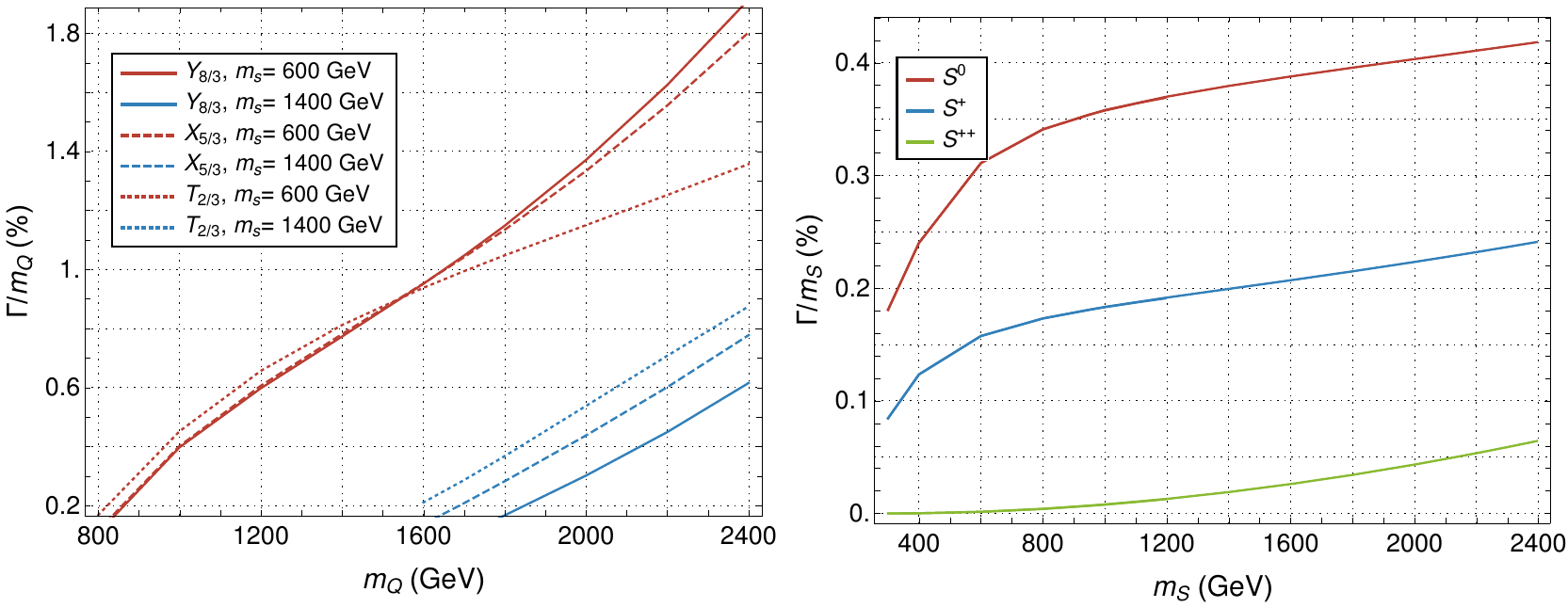} %\hspace{10pt}
\end{center}    	
    \caption{\sf\it Total decay widths of the VLQs (left) and scalars (right) as functions of their respective masses. The decay widths of the scalars are shown for $m_Q=1700$ GeV, however, they are approximately independent of $m_Q$.}
    \label{fig:totwidth}
\end{figure}

In \cref{fig:totwidth} the total decay widths of the VLQs and the scalars are displayed for some representative mass points. In the entire mass range considered in our analysis, the VLQs have $\Gamma/m_Q<2\%$ and the scalars have $\Gamma/m_S<1\%$, thus validating the use of narrow width approximation. 

% % % % % % % % % % % % % % % % % % % % % % % % % % % % % % % % % % % % % %
\section{LHC constraints}
\label{sec:LHCconstraints}
% % % % % % % % % % % % % % % % % % % % % % % % % % % % % % % % % % % % % %

\begin{figure}[h!]
	\centering
	\includegraphics[trim={2.5cm 22.5cm 1.8cm 2.3cm},clip,scale=1]{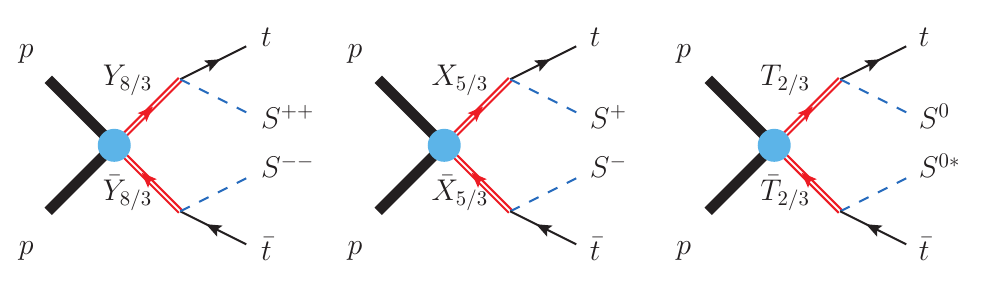}
	\caption{\sf\it Feynman diagrams contributing to the pair production of VLQs at the LHC.}
	\label{fig:pair_production}
\end{figure}

To obtain the current bounds on the masses of the VLQs and scalars of the model, we have performed a recast of LHC data. For this purpose, we have implemented the model in \texttt{FeynRules} \cite{Alloul:2013bka} by extending an existing model \cite{Banerjee:2022xmu} and, generated a UFO \cite{Degrande:2011ua} model file with four massless quarks, suitable for simulation at NLO in QCD, using \texttt{FeynArts} \cite{Hahn:2000kx} and \texttt{NLOCT} \cite{Degrande:2014vpa}. We have focused exclusively on pair production of the VLQs at the LHC with subsequent decays into the scalar triplet, see \cref{fig:pair_production}, and simulated separately the production of the three VLQs, assuming that interferences between signal topologies from different VLQs are negligible. This is reasonable due to the different charges of the VLQs and their largely different decay patterns. The advantage of considering pair production is that the cross-section is model-independent and only varies with the mass of the VLQs.

\begin{figure}[h!]
	\centering
	\includegraphics[width=.5\textwidth]{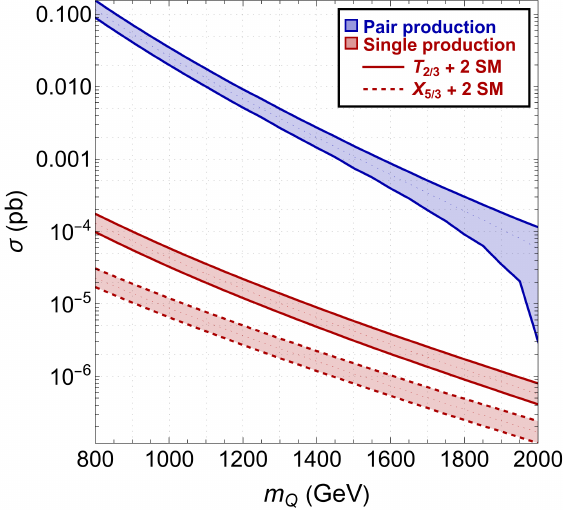}
	\caption{\sf\it Cross-sections at LO for pair production and single production of $T_{2/3}$ and $X_{5/3}$. Systematic uncertainties are represented by the bands around the central values.}
	\label{fig:pair_vs_single}
\end{figure}

As shown in \cref{fig:pair_vs_single}, single production of $T_{2/3}$ and $X_{5/3}$ have negligible cross-sections in comparison to the pair production process due to suppressed couplings of the VLQs with the $W$ and $Z$ bosons, thus they have not been considered in our analysis. Further, $X_{8/3}$ cannot be produced singly without the propagation of either other VLQs or new scalars. For the purpose of showing the relative importance of single production with respect to pair production, only LO results have been obtained. Clearly, going to NLO precision would not compensate the large difference.

Exclusion bounds on the masses of the VLQ and the scalar triplets have been obtained by recasting the ATLAS and  CMS searches available in the \texttt{MadAnalysis5}~\cite{Conte:2012fm,Conte:2014zja,Dumont:2014tja} public analysis database (PAD). The event generation has been done through \texttt{MG5\_aMC}~\cite{Alwall:2014hca}, while hadronization and parton showering have been performed through \texttt{Pythia 8}~\cite{Sjostrand:2014zea}. 
The decay chains have been simulated within \texttt{MG5\_aMC} preserving spin correlations. We consider the possible decays (up to 3-body decays) of each of the VLQs as shown in \cref{decays}. To optimize the use of computational resources we only consider the decays with appreciable branching ratios (BR $\geq1\%$) in the range of parameter space we are interested in. Since BSM scalars are lighter than the VLQs, they only decay into SM states, see \cref{fig:pngb_br}. 

Due to the complexity of the decay chains, simulations have been performed at LO in QCD using the \texttt{NNPDF~3.1} parton distribution functions (PDFs)~\cite{NNPDF:2017mvq}, taken from the {\sc LHAPDF 6} library~\cite{Buckley:2014ana}, but a $K$-factor is associated to the cross-sections of the simulated samples, calculated at the NNLO+NNLL accuracy using \texttt{Top++}\cite{Czakon:2011xx} with the \texttt{NNPDF4.0}~\cite{NNPDF:2021njg} PDFs.

The mass range of the VLQs has been taken between 800 GeV and 2 TeV. The minumum value has been chosen well below current bounds because in principle, due to the exotic decays, current VLQ bounds based on purely SM decays have to be rescaled~\cite{Banerjee:2022xmu}, and multiple studies have shown that this might imply a reduction of the bounds. For the scalar triplet, we started from $m_S\geq 400$ GeV to avoid bounds from the direct search of scalars through their pair production by Drell-Yan~\cite{Banerjee:2022xmu}. 

\begin{figure}[h!]
\begin{center}
    	\includegraphics[width=0.5\linewidth]{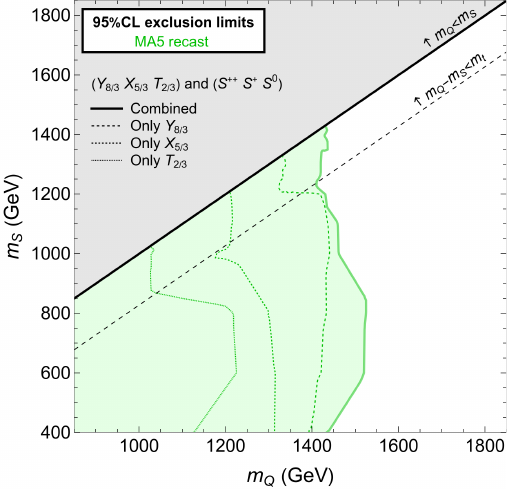} \hspace{10pt}
\end{center}    	
    \caption{\sf\it Exclusion limits for the individual pair productions of the three VLQs, $T_{2/3}$, $X_{5/3}$ and $Y_{8/3}$, and for the sum of their signals.}
    \label{fig:exclusion_summary}
\end{figure}

Individual exclusion limits and the exclusion for the sum of the signals coming from the three VLQs are shown in \cref{fig:exclusion_summary}.
The $T_{2/3}$ is excluded up to 1.2 TeV in RL, and the limit drops to around 1 TeV in RS, where SM decays dominate; the $X_{5/3}$ has higher bounds, around 1.3 TeV in RL and 1.2 TeV in RS; the $Y_{8/3}$ has the strongest bounds, which reach and slightly surpass 1.4 TeV in RL, and reduce to around 1.35 TeV in RS.

However, it is crucial to note that the actual limit is determined by the sum of the signals contributed by all three VLQs present in the model. This reaches 1.5 TeV in RL and reduces to around 1.4 TeV in RS. The bound is clearly dominated by the $Y_{8/3}$ contribution, but for some mass configurations in both RL and RS it becomes significantly higher than the bound on the mass of $Y_{8/3}$ when considered individually, signalling that the contributions of the other two VLQs are not negligible.

This is an important aspect, which will be further investigated in \cref{sec:HL_LHC_prospect}: in realistic theoretical models there are usually more than one VLQs, potentially close in mass, {\em e.g.} arising from a $SU(2)_L$ multiplet. They can produce a significant number of signal events even if their masses are too high for being observed individually. The observation of excesses corresponding to high invariant masses can then be interpreted as coming from a combination of signal events of multiple VLQs with similar mass.

To conclude this section, we notice that the bounds are determined predominantly by a few signal regions (SRs) of two CMS searches for RL~\cite{CMS:2019rvj} and RS~\cite{CMS:2019lwf}, respectively. The cuts characterising the respective signal regions are described in \cref{tab:recast}.\footnote{Consult \cite{CMS:2019rvj,CMS:2019lwf} for more details about the baseline selection on the trigger requirements, the momenta of the leptons etc.} A detailed description of which search and SR performs better for each point of our scan is provided in \cref{app:recastSR}.

\begin{table}[h!]
    \centering
  %  \begin{small}
    \def\arraystretch{2.0}
    \begin{tabular}{clccccclccc}
        \toprule\toprule 
        Searches & Kinematics & SR & $N_l$ & $N_{ \sf\rm OSSF}$& $N_b$ & $N_j$ & Limits on $m_Q$\\
        \midrule
       ~\\[-28pt]
       \multirow{2}{*}{\cite{CMS:2019rvj} } & \multirow{2}{*}{\parbox{2.5cm}{$H^j_T>300$ GeV \\ $\slashed p_T>50$ GeV} } 
       % &  SR1 & 2 & -- & 2 & 6 & \parbox{3.5cm}{$m_Q<m_S+m_t$ \\ $m_Q\lesssim 900$ GeV}\\
       % \cline{3-8}
        %SR2 & 2 & 2 & 7\\
        %SR3 & 2 & 2 & $\geq 8$\\
       & SR7 & 2~\text{same-sign} & -- & 3 & $\geq 8$ & \parbox{2.5cm}{$m_Q<m_S+m_t$}\\[1pt]
       \cline{3-8}
       & & SR8 & 2~\text{same-sign} & -- & $\geq4$ & $\geq 5$ & \parbox{2.5cm}{$m_Q<m_S+m_t$ \\ $m_Q\lesssim 1700$ GeV}\\[3pt]
       \midrule
       ~\\[-23pt]
      \cite{CMS:2019lwf} & \parbox{3.cm}{$M_{\sf\rm OSSF}>106$ GeV \\ $L_T+\slashed E_T \in$ \\$[875,1000]$ GeV}  & 3L above-Z & 3 & 1 & -- & -- & \parbox{2.5cm}{$m_Q>m_S+m_t$}\\[8pt]
        \bottomrule\bottomrule
    \end{tabular}
  %  \end{small}
    \caption{\sf\it Most sensitive signal regions for the recast. OSSF stands for opposite-sign same-flavour leptons.}
    \label{tab:recast}
\end{table}
Clearly, a dedicated recast of more recent searches targeting these objects would probably improve the result, but this is beyond the scope of the current analysis. It is important to stress that, in RL, the largest region of the parameter space, a search targeting four tops and mostly based on the multiplicity of final state particles (number of same-sign leptons, of jets and of $b$-jets) with minimal kinematic cuts is sensitive to new physics which lead to final states rich in these objects, such as the case of the scenario at hand. In \cref{sec:HL_LHC_prospect} we will follow a similar strategy, relying on minimal selection criteria together with strong cuts over global variables which can be sensitive to new physics at high masses in the HL-LHC phase, and further evaluate how it performs with respect to the best SRs of the recast.

\section{Prospects for HL-LHC}
\label{sec:HL_LHC_prospect}

The model we are considering has two crucial features, which can be exploited to design a dedicated analysis:
\begin{enumerate}
\item the presence of a VLQ with charge $8/3$, which can generate final states with multiple same-sign leptons and can be used as a smoking gun of this scenario;
\item the simultaneous presence of multiple VLQs and scalars, arising from triplets and in the same mass range: the former, as also mentioned in \cref{sec:LHCconstraints}, all contribute to produce signal events in the same invariant mass region, while the latter introduce a large range of decay channels which can lead to final states with high multiplicity of jets, b-jets and leptons.
\end{enumerate}

To analyze the HL-LHC prospect we consider two benchmark mass points, one in RL and the other in RS:
\begin{equation}{\setlength{\arraycolsep}{2pt}\left\{\begin{array}{lll} \text{BPL:} & m_Q=1700~{\rm GeV},~m_S=600~{\rm GeV},\\[-3pt] \text{BPS:} & m_Q=1700~{\rm GeV},~m_S=1600~{\rm GeV}. \end{array}\right.}\;
\label{eq:BPs}
\end{equation}
These points have been chosen such that the combination of the three VLQ signals is not excluded by current bounds but their mass is close enough be in the reach (exclusion or discovery) of the HL-LHC.
For the BPL, the most relevant final states in order of importance  are listed in \cref{tab:dominant_channels} for the three VLQs of different charges.
\iffalse
\begin{align*}
    Y_{8/3} + \bar{Y}_{8/3} & \to 3 W^+ + 3 W^- + 6 b ~\quad \qquad > 90\% \\ \\
    X_{5/3} + \bar{X}_{5/3} & \to 2 W^+ + 2 W^- + 6 b ~\quad \qquad\sim 60\% \\
                            & \to 3 W^+ + 3 W^- + 6 b ~\quad \qquad\sim 16\% \\
                            & \to 2 W^+ + 2 W^- + 4 b ~\quad \qquad\sim 12\% \\
                            & \to 2 W^+ + 2 W^- + 4 b ~\quad \qquad\sim 12\% \\ \\    
    T_{2/3} + \bar{T}_{2/3} & \to 1 W^+ + 1 W^- + 6 b ~\quad \qquad \sim 22\% \\
                            & \to 3 W^+ + 1 W^- + 6 b ~\quad \qquad \sim 15\% \\
                            & \to 1 W^+ + 3 W^- + 6 b ~\quad \qquad \sim 15\% \\
                            & \to 3 W^+ + 3 W^- + 6 b ~\quad \qquad \sim 10\% \\
                            & \to 1 W^+ + 1 W^- + 4 b  + X_{\slashed b}\quad \sim 6\% \\
\end{align*}
\fi
%\multicolumn{3}{c}{\parbox{2.4cm}{Final state multiplicities}}
\begin{table}[h!]
    \centering
    \def\arraystretch{1.3}
    \begin{tabular}{cccclc}
    \toprule\toprule
    VLQ pair & \multicolumn{3}{c}{\parbox{3.2cm}{No. of b-jets, $W^\pm$ \\ from VLQ decay}} & Contributing decays & Product of BRs \\
    & $N_b$ & $N_{W^+}$ & $N_{W^-}$ & & \\
    \midrule
    $Y_{8/3} + \Bar{Y}_{8/3}$ & 6 & 3 & 3 & \parbox{4cm}{$Y_{8/3}\to t + S^{++}$ \\ $~~~~~~\to t + W^+ + S^{+}$ \\ $~~~~~~\to b + W^+ + S^{++}$} & $>92\%$ \\[10pt]
    \midrule 
    \multirow{3}{*}{$X_{5/3} + \Bar{X}_{5/3}$} & 6 & 2 & 2 & $X_{5/3}\to t + S^{+}$ & $>50\%$ \\[2pt]
    \cline{2-6}
    ~\\[-13pt]
    & 6 & 3 & 3 & \parbox{4cm}{$X_{5/3}\to t + S^{+}$ \\ $~~~~~~~\to t + W^- + S^{++}$} & $>18\%$ \\[5pt]
    \cline{2-6}
    & 4 & 2 & 2 & \parbox{4cm}{~\\$X_{5/3}\to t + S^{+}$ \\ $~~~~~~~\to t + W^+$} & $>12\%$ \\
    \midrule
    \multirow{6}{*}{$T_{2/3} + \Bar{T}_{2/3}$} & 6 & 3 & 3 & $T_{2/3} \to t + (S^0 \to t\Bar{t})$ & $>9\%$ \\[5pt]
    \cline{2-6}
    & 6 & $<3$ & $<3$ & \parbox{4cm}{~\\$T_{2/3} \to t + S^0$ \\ $~~~~~~\to t + (Z \to b\Bar{b})$ \\ $~~~~~~\to t + (h \to b\Bar{b})$} & $>53\%$ \\[15pt]
    \cline{2-6}
    & 4 & $\geq 1$ & $\geq 1$ & \parbox{4cm}{~\\$T_{2/3} \to t + S^0$ \\ $~~~~~~\to t + Z$ \\ $~~~~~~\to t + h$} & $>11\%$ \\
    \bottomrule\bottomrule
    \end{tabular}
    \caption{\sf\it Minimum value of the product of BRs of the VLQ pairs for $m_Q=1700$ GeV and $m_S=600$ GeV, categorized according to final states with different multiplicities of b-jets and $W$ bosons. 
    }
    \label{tab:dominant_channels}
\end{table}

The analysis uses simulated signal and background events. The parton-level Monte Carlo (MC) events are generated using \texttt{MG5\_aMC} at $\sqrt{s}=13$ TeV, and showered using \texttt{Pythia 8}~\cite{Sjostrand:2014zea}. These events are further passed through a fast-detector simulation using \texttt{Delphes~3}~\cite{deFavereau:2013fsa}. All simulations use the default ATLAS detector card, with a reduction of the radius parameter for the lepton and photon isolation from $0.5$ to $0.2$. As the signal topologies consist of final states with large multiplicities, loosening the isolation requirement increases the sensitivity to the signal of interest.

We simulate 100K events for each VLQ type and for each benchmark point.
Several SM processes are relevant as backgrounds for this analysis. Following \cite{Han:2018hcu}, to estimate the number of background events in the signal region of interest, we simulate $4t$, $t\bar{t}V+\leq 2j$ (where $V$ denotes $W$ and $Z$ bosons), $t\bar{t}+ \leq 3j$, $t\bar{t}b\bar{b}$, and $VVV$ processes. Of these, 200K events are simulated for $t\bar{t}V$, 500K events for $t\bar{t}+ \leq 3j$, and 100K events for the rest. We further require the $t\bar{t}V+\leq 2j$, $t\bar{t}+ \leq 3j$ events to satisfy $\sqrt{\hat{s}} > 1200$~GeV in order to better model the high-$H_T$ tail.

For the background simulations, the additional jets from initial- and final-state QCD radiation are included at both matrix-element and parton-shower level; double counting is removed through the MLM jet-merging algorithm implemented in \texttt{MG5\_aMC}~\cite{Mangano:2006rw} with parameters \verb#xqcut#=30 and \verb#qcut#=45. For HL-LHC projections, both signal and background simulations have been done at LO using the NNPDF~3.1 PDFs.

\subsection{Signal regions}

The signal processes can yield final states with number of leptons ranging from 0 to 6. An interesting direction would be to try to identify the exotic charge of $Y_{8/3}$ through the distributions of same-sign leptons in final states with large lepton multiplicity. However, requiring too many leptons in the final state in the VLQ mass range still allowed by the recast bounds dramatically reduces the signal yield owing to the low branching ratio of $W$ bosons into leptons. Therefore, designing a SR to discriminate a $Y_{8/3}$ seems quite challenging. On the other hand, a fully hadronic final state results in a high background yield. 

Our strategy will be therefore to impose minimal selection cuts on the number of leptons and on jets and $b$-jets, and exploit the fact that we are targeting VLQs with masses above 1.5 TeV by posing strong cuts on observables related to the total energy of final state.

\begin{figure}[h!]
    \centering
         \includegraphics[width=0.495\textwidth]{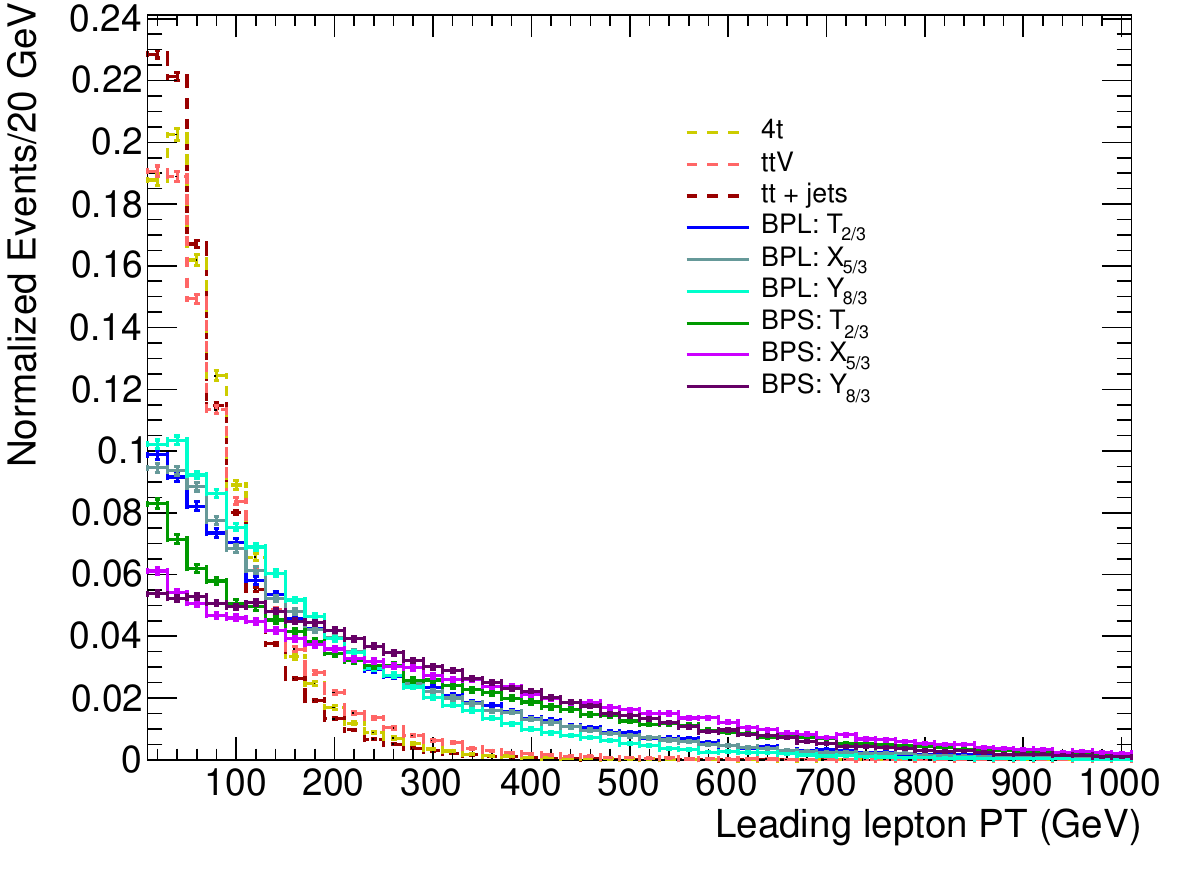}
         \includegraphics[width=0.495\textwidth]{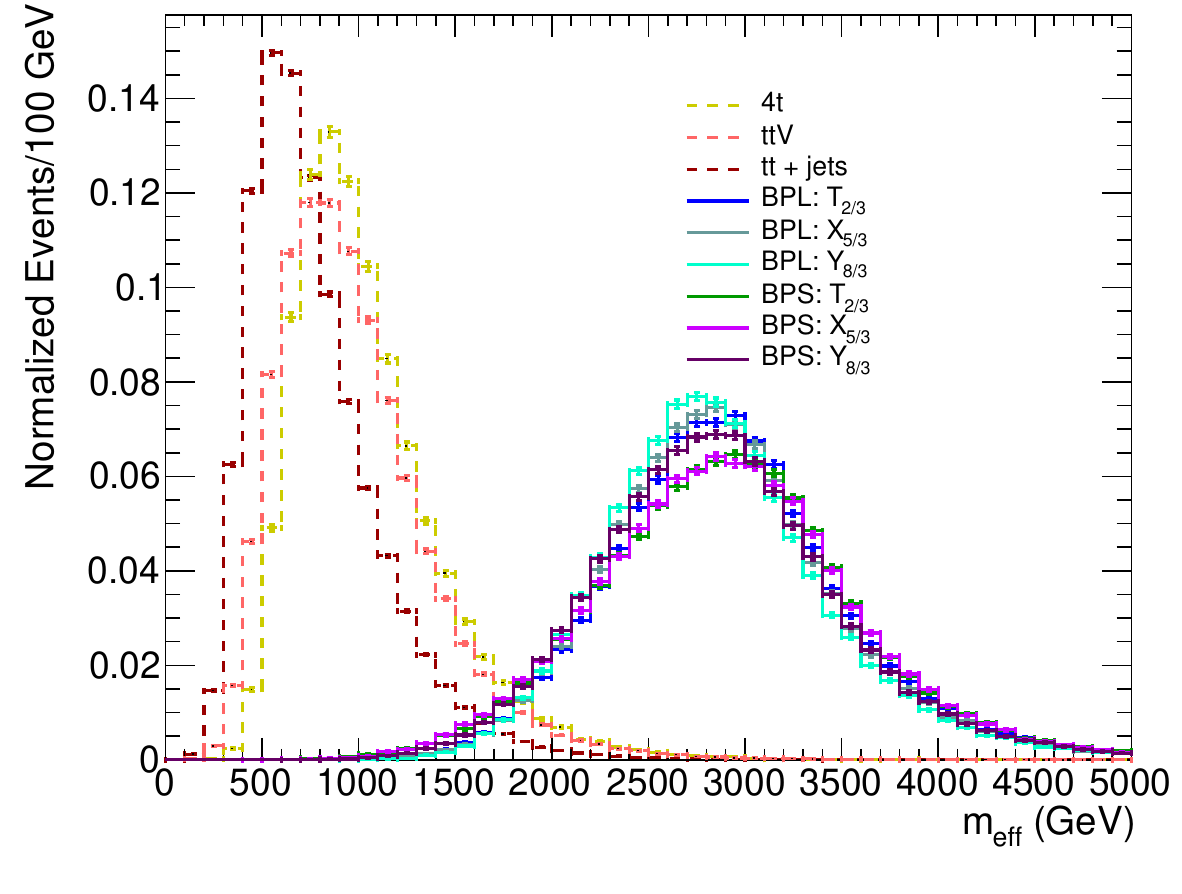}
    \caption{\sf\it Signal and background distributions of the leading $b-$jet $p_T$ and the effective mass without any selections applied}
    \label{fig:SigVsBkg}
\end{figure}

We define two SRs, summarised in \cref{tab:SRs}, designed to maximize signal sensitivity for the two BPs defined in \cref{eq:BPs}: the first, labelled as SRL, targets BPL, while the second, labelled SRS, targets BPS. 
\begin{table}[h!]
\centering
\begin{tabular}{cccccc}
\toprule\toprule 
SR & $N_{\rm SSL}$ & $N_j$ & $N_b$ & $p_T(l_0)$ & $m_{\rm eff}$ \\
\midrule
SRL & \multirow{2}{*}{$\geq 1$} & \multirow{2}{*}{$\geq 3$} & $\geq 2$ & $-$ & \multirow{2}{*}{$\geq 2100$ GeV or $\geq 2300$ GeV} \\
\cmidrule(r{2pt}){1-1} \cmidrule(l{2pt}){4-5}
SRS & & & $\geq 1$ & $\geq 170$ GeV \\
\bottomrule\bottomrule
\end{tabular}
\caption{\sf\it Signal regions for the HL-LHC analysis.}
\label{tab:SRs} 
\end{table}
Given the number of $b$-jets and $W$ bosons for the signal (as shown in \cref{tab:dominant_channels}), both the SRs are characterised by the presence of at least one pair of same-sign leptons and at least three jets. Due to the large number of $b$-jets in the large mass gap region of BPL, the signal region SRL also requires at least two $b$-jets. The number of $b$-jets in BPS is limited due to the fact that all VLQs almost exclusively decay directly into SM particles, so that $b$-jets could only come from the decay of top quarks, only one per branch, or $Z$ and $h$ bosons. Therefore the majority of signal events for BPS will contain two $b$-jets, further subject to tagging efficiency. To avoid depleting the signal through a very strong selection, SRS will therefore require only one $b$-jet in the final state. Furthermore, to exploit the shorter decay chains of BPS, in SRS we impose a cut on the transverse momentum of the leading lepton, $p_T(l_0)\geq 170$ GeV. Finally, to target the high VLQ masses, we impose cuts on the effective mass, $m_{\rm eff}$, defined as the scalar sum of all visible objects in the detector and the missing transverse momentum, required to be larger than 2100 GeV or 2300 GeV in both SRs. This variable is known to be efficient in discriminating signal from background in searches for BSM particles with high mass (for example, see \cite{ATLAS:2022tla}). In \cref{fig:SigVsBkg} the distributions of $p_T(l_0)$ and $m_{\rm eff}$ are shown with no selection applied. 

With these SRs the yields of the $t\bar{t}b\bar{b}$, $VVV$ and $t\bar t + \leq 3j$ backgrounds are negligible, and will not be used in the statistical analysis. The reason for testing two slightly different cuts on $m_{\rm eff}$ is that while increasing the value (up to $\sim 2500$ GeV) we are effectively cutting backgrounds while preserving most of the signal, above 2300 GeV the number of MC background events almost entirely vanishes (see \cref{tab:xsec_eff}). Above 2300 GeV we become effectively limited by MC statistics, and cutting too strongly would significantly deplete the MC statistics introducing large uncertainties to our analysis. 
\begin{table}[h!]
    \centering
    \def\arraystretch{1.3}
    \begin{tabular}{ccccccccccc}
    \toprule\toprule
   SR & Backgrounds & $\sigma$ [fb] & & $\epsilon(m_{\rm eff}>2100$ GeV) & & $\epsilon(m_{\rm eff}>2300$ GeV) \\
    \midrule\midrule
     SRL & \multirow{2}{*}{$\begin{array}{c} t\Bar{t}V+\leq 2j\\[-3pt] \text{\footnotesize (with $\sqrt{\hat s}\geq 1200~$ GeV)} \end{array}$} & \multirow{2}{*}{838} & & $1.20\times 10^{-4}$ & & $5.24\times 10^{-5}$\\
     SRS & & &  & $9.43\times 10^{-5}$ & & $3.24\times 10^{-5}$\\
     \midrule
     SRL & \multirow{2}{*}{$4t$} & \multirow{2}{*}{5.32} &  & $3.20\times 10^{-4}$ & & $1.70\times 10^{-4}$\\ 
     SRS & & &  & $2.00\times 10^{-4}$ & & $1.20\times 10^{-4}$ \\ 
    \midrule\midrule 
    BP/SR & Signal & $\sigma$ [fb] & & $\epsilon(m_{\rm eff}>2100$ GeV) & & $\epsilon(m_{\rm eff}>2300$ GeV) \\
    \midrule\midrule
    \multirow{3}{*}{BPL/SRL} & $Y_{8/3}$ pair & 3.07 & & 0.092 & & 0.079\\
    & $X_{5/3}$ pair & 3.21 & & 0.051 & & 0.044 \\
    & $T_{2/3}$ pair & 3.19 & & 0.030 & & 0.026\\
    \midrule
    \multirow{3}{*}{BPS/SRS} & $Y_{8/3}$ pair & 3.15 & & 0.088 & & 0.077\\
    & $X_{5/3}$ pair & 3.19 & & 0.035 & & 0.031\\
    & $T_{2/3}$ pair & 3.16 & & 0.025 & & 0.022\\
    \bottomrule\bottomrule
    \end{tabular}
    \caption{\sf\it Signal and background cross-sections ($\sigma$) and efficiencies ($\epsilon$).}
    \label{tab:xsec_eff}
\end{table}

\subsection{Numerical results and interpretation}
\label{sec:resultsinterpretation}

The signal and background yields after applying the signal region cuts and assuming a dataset corresponding to the the nominal luminosity of $3$ ab$^{-1}$ are shown in Table~\ref{tab:xsec_eff}. The SRs could be further optimized using cuts on other variables but this optimization is not performed for this analysis due to the statistical limitation of our simulations.

The expected discovery and exclusion significance are calculated using the number of signal ($S$), and background ($B$) events along with the systematic uncertainties in the background ($\sigma_B$), following the expressions \cite{Kumar:2015tna}
\begin{align}
    Z_{\rm disc} = \sqrt{2}\left[(S+B)\ln\left(\frac{(S+B)(B+\sigma_B^2)}{B^2+(S+B)\sigma_B^2}\right)-\frac{B^2}{\sigma_B^2}\ln\left(1+\frac{\sigma_B^2 S}{B(B+\sigma_B^2)}\right)\right]^{1/2}\;,
    \label{asimov_disc}
\end{align}
for the discovery reach and
\begin{align}
    Z_{\rm exc} = \left[2\left\{S-B\ln\left(\frac{B+S+x}{2B}\right)-\frac{B^2}{\sigma_B^2}\ln\left(\frac{B-S+x}{2B}\right)\right\}-(B+S-x)\left(1+\frac{B}{\sigma_B^2}\right)\right]^{1/2},
    \label{asimov_exc}
\end{align}
for the exclusion limit, where: 
\begin{align}
    x\equiv \sqrt{(S+B)^2-\frac{4S B \sigma_B^2}{B+\sigma_B^2}}.
\end{align}

\begin{figure}[h!]
\begin{center}
    	\includegraphics[width=\textwidth]{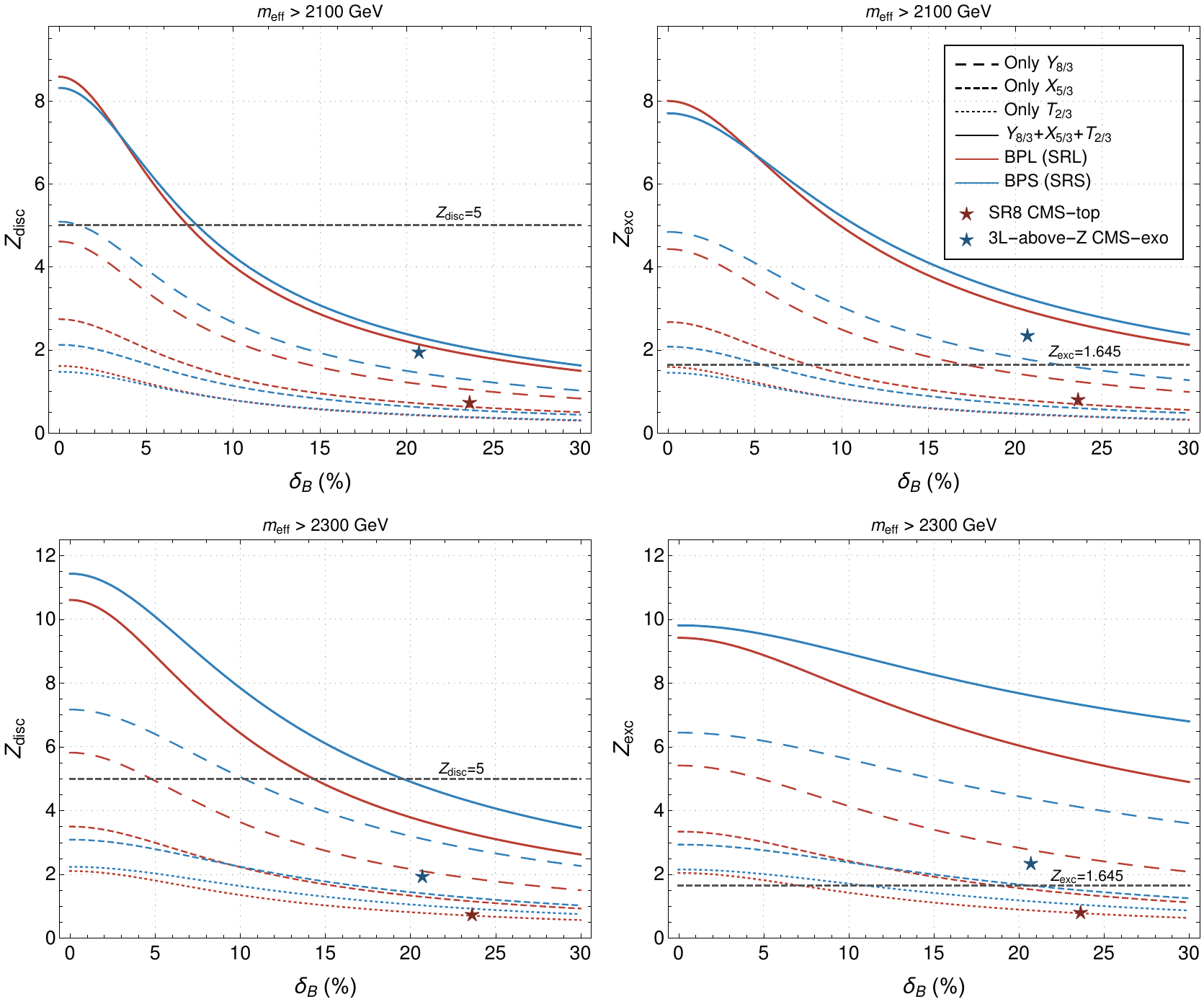} \hspace{10pt}
\end{center}    	
    \caption{\sf\it Expected significance for discovery (left) and exclusion (right) at the HL-LHC are shown in our proposed SRs for two kinematic cuts, $m_{\rm eff}>$2100 (top) and 2300 (bottom) GeV, respectively. The asterisk marks denote the projected HL-LHC significance for the SRs shown in \cref{tab:recast}, namely, SR8 from \cite{CMS:2019rvj} (red) and 3L-above-Z from \cite{CMS:2019lwf} (blue).}
    \label{fig:projection_summary}
\end{figure}

To show realistic results, background systematics have to be estimated. While we have simulated the main sources of background, our analysis relies on a fast detector simulation and completely neglects further data-driven contributions which can modify the overall background acceptances. Instead of estimating a systematic uncertainty, we prefer to parametrise it and show our results as function of the relative systematic uncertainty $\delta_B=\sigma_B/B$. In \cref{fig:projection_summary} we show the expected exclusion and discovery significance for the three individual VLQ pair production processes and for their combination, for both BPL and BPS, in a range of $\delta_B$ from 0 to 30\%. In the same plot we also show the projected significances for the best SRs of the recast, assuming their systematic uncertainties do not depend on the luminosity.\footnote{In order to compare results consistently, projections from the recast have been obtained using LO cross-sections for the signal and not the NNLO+NNLL values used to obtain the bounds.} Results are shown for the two different $m_{\rm eff}$ cuts, as described in \cref{tab:SRs}. 

A number of conclusions can be inferred:
\begin{itemize}
\item If VLQs are considered individually, they cannot be discovered ($Z_{\rm disc}\geq5$) at HL-LHC with the SRs designed in this analysis, with the exception of $Y_{8/3}$, and only if systematics can be pushed to very low (probably too optimistic) values, less than 10\% for BPS and 5\% for BPL. This is valid only when imposing the strongest $m_{\rm eff}$ cut above 2300 GeV.
\item On the other hand, the combined signal $T_{2/3}+X_{5/3}+Y_{8/3}$ can be discovered with systematic uncertainties up to 15\% (20\%) for BPL (BPS) when the $m_{\rm eff}\geq2300$ GeV cut is applied.
\item For exclusion limits ($Z_{\rm exc}\geq1.645$~\cite{Kumar:2015tna}), it is possible to see that while the combined signal can always be excluded, for both BPL and BPS and even for large systematic uncertainties, individual VLQs can only be excluded if systematics can be reduced below certain values, depending on the VLQ and on the $m_{\rm eff}$ cut. Exclusion significances can be estimated for higher VLQ masses: considering 20\% systematics and (conservatively) assuming same signal efficiencies for higher masses, a $Z=1.645$ exclusion is obtained around $\sim1.9$ TeV using the higher $m_{\rm eff}\geq2300$ GeV cut.
\item The performance of the two SRs designed for BPL and BPS is better than the projected values of the best SRs of the recast for equal systematic uncertainty. It is interesting to notice that the projected exclusion limit for the best recast SR for BPL cannot exclude the combined signal. This sizably stronger result confirms that the application of global variable cuts can indeed improve the sensitivity to new physics at high mass scales. This is especially true when considering processes where mass differences in the chain decays are not large, and therefore a search strategy cannot rely on boosted objects.
    
\end{itemize}
We remind that these results have been obtained under the assumption that the three VLQs are degenerate in mass, and same for the scalars. We have verified (\cref{app:nondegenerate}) for BPL that lifting the degeneracies by 50 GeV does not significantly impact our conclusions.

% % % % % % % % % % % % % % % % % % % % % % % % % % % % % % % % % % % % % %
\section{Conclusions}
\label{conc}

We have investigated the phenomenology at the LHC and HL-LHC of a non-minimal scenario where the SM is extended by a VLQ triplet with $Y=5/3$ and a complex scalar triplet with $Y=1$, theoretically motivated by models of partial compositeness involving a coset $SU(5)/SO(5)$. 
This scenario features a vector-like top partner and two vector-like quarks with exotic charges, 5/3 and 8/3; its scalar sector contains one new neutral scalar which does not acquire a vacuum expectation value, a charged scalar and a doubly-charged scalar.

The values of the couplings of new particles between themselves and with SM states are in principle free parameters, but for our analysis they have been chosen in accordance with the partial compositeness scenario. In particular, owing to the $\mathcal{O}(1)$ couplings between the VLQs, the complex scalar triplet and a right-handed top quark, we show that above the kinematical threshold $m_Q = m_S+m_t$ the leading decay channels of the VLQs in this model are $Y_{8/3}\to tS^{++}$, $X_{5/3}\to tS^{+}$, and $T_{2/3}\to tS^{0}$. Affinity of the BSM scalars towards the third generation quarks, again motivated by partial compositeness, dictates the decay widths and BRs of the scalars. Notably, the doubly charged scalar exhibits a unique 3-body decay channel $S^{++}\to t\Bar{b}W^+$ with 100\% BR. 

Assuming a degenerate spectra for the components of the VLQ triplet, and similarly for the scalar triplet, we establish exclusion limits in the $m_Q$ vs. $m_S$ plane by recasting a set of LHC experimental searches using data from Run 2. We found that the most sensitive signal regions exclude at 95\% CL a VLQ mass up to around 1.5 TeV in a region where the mass splitting between the VLQs and the new scalars is larger than the mass of the top quark. The exclusion limit reduces by around 100 GeV in a small mass splitting region where $m_Q-m_S\lesssim m_t$.

An important feature of our analysis is that the $\sim1.5$ TeV exclusion limit quoted above is on the {\it combined} signal coming from pair-production of all the VLQs of the triplet, such that all signal events from the production of the individual VLQs contribute to the overall exclusion. Indeed, the exclusion limits of each VLQs considered individually are much lower, up to 1.2 TeV for $T_{2/3}$, 1.3 TeV for $X_{5/3}$ and 1.4 TeV for $Y_{8/3}$ in the large mass splitting region, and analogous reduction in the small splitting region. 

Motivated by the possibility of combining signal events from multiple VLQs to extend the reach of future searches, we further explore the discovery prospect of this model at the HL-LHC with nominal integrated luminosity of 3 ab$^{-1}$. We have designed two signal regions to target final states corresponding to large and small mass gaps between the VLQ and the scalars. The main feature of our strategy consists in imposing a strong cut on the effective mass ($m_{\rm eff}$) of the final state, defined as the scalar sum of the transverse momenta of all visible particles in the final state and missing transverse momentum. Cuts on global variables such as the effective mass are known to be sensitive to new physics at very high energy scales, therefore if the overall effective cross-section of the signal (individual contributions modulated by experimental acceptances) is high enough, such cuts can be very powerful for probing signal events from multiple new particles at similar mass scales, which would otherwise be out of reach individually. 

We have found that if systematic uncertainty of the background can be reduced below $\sim 20\%$, our analysis strategy can lead to a 5$\sigma$ discovery of signals coming from the sum of the three VLQs in the triplet with $m_Q=1700$ GeV, for two different choices of $m_S$ (600 GeV and 1600 GeV), outperforming the projected discovery reaches of the best signal regions of the recast for the same benchmark points. Individual VLQs, on the other hand, cannot be discovered with our SRs, with the exception of $Y_{8/3}$ in the very optimistic hypothesis that systematics can be pushed below 5\%. The exclusion reach of the combined signal can approach a VLQ mass up to 2 TeV. 

Our analysis focuses on a specific scenario involving triplets, but our results have a much broader reach: theoretical scenarios usually predict VLQs and new scalars in one or more multiplets, the components of which may have masses in a similar range, and even if their values lay above current limits, their combined effect might increase signal yields at high mass scales. In our scenario, the difference between the electric charges of the VLQs does not allow for strong signal-signal interferences, but if multiple VLQs with same charge and similar masses are present, these effects can further significantly increase signal yields.
The quest for VLQs at the LHC is therefore still open.
Even if VLQs are too heavy to be observed individually, excesses can still appear due to a combination of signals around the same energy scale. Exploring regions towards 2 TeV with VLQ pair production is therefore feasible, theoretically well justified, and has the potential not only to significantly improve current mass bounds, but also to lead to new discoveries.

% % % % % % % % % % % % % % % % % % % % % % % % % % % % % % % % % % % % % %

\section*{Acknowledgements}

The authors thank Rikard Enberg and Gabriele Ferretti for collaboration during the initial stages of the work. The work is supported by the Knut and Alice Wallenberg foundation (Grant KAW 2017.0100, SHIFT project). LP's work is supported by ICSC – Centro Nazionale di Ricerca in High Performance Computing, Big Data and Quantum Computing, funded by European Union – NextGenerationEU. LP acknowledges the use of the IRIDIS HPC Facility at the University of Southampton.

% % % % % % % % % % % % % % % % % % % % % % % % % % % % % % % % % % % % % %
\appendix
% % % % % % % % % % % % % % % % % % % % % % % % % % % % % % % % % % % % % %

% % % % % % % % % % % % % % % % % % % % % % % % % % % % % % % % % % % % % %
\section{Effective field theory construction}
\label{EFT}
% % % % % % % % % % % % % % % % % % % % % % % % % % % % % % % % % % % % % %

One can construct an EFT including operators up to dimension-5 by extending the SM with a $Q$ and $S$ multiplets as
\begin{equation}
\mathcal{L}=\mathcal{L}_{\rm SM}+\mathcal{L}_{\rm NP}^{d\leq 4}+\mathcal{L}_{\rm NP}^{d=5},    
\end{equation}
where
\begin{align}
\label{EFT_1}
\mathcal{L}_{\rm NP}^{d\leq 4} & = |D_\mu S|^2 - m_S^2|S|^2 + \bar{Q}\left(i\slashed D - m_Q\right) Q + \lambda_R \bar{Q}_L S t_R + {\rm h.c.} \\
\mathcal{L}_{\rm NP}^{d=5} & = \frac{\tilde{y}_t}{\Lambda}\bar{q}_L S^\dagger H t_R + \frac{\tilde{y}_b}{\Lambda}\bar{q}_L S H^c b_R + \frac{\tilde{\lambda}_1}{\Lambda}H^\dagger i\tau^2\bar{Q}_L H^* t_R + \frac{\tilde{\lambda}_2}{\Lambda}\bar{q}_L S^\dagger Q_R H^c + {\rm h.c.}
\label{EFT_2}
\end{align}
We neglect interactions between $S$ and the Higgs doublet $H$, since they play no role for the purpose of this work. The mass matrix for $(t, T_{2/3})$ below the electroweak symmetry breaking scale is given by
\begin{equation}
\mathcal{M}=\left(\begin{array}{cc}
\frac{y_t v}{\sqrt{2}}     &  0\\
-\frac{\tilde{\lambda}_1 v^2}{2\Lambda}     & m_Q
\end{array}\right).    
\end{equation}
The masses of top quark and $T_{2/3}$ at the leading order in $v/\Lambda$ are given by
\begin{equation}
m_t = \frac{y_t v}{\sqrt{2}}\left(1- \frac{\tilde{\lambda}_1^2}{8}\frac{v^4}{\Lambda^2m_Q^2}\right), \quad m_{T_{2/3}} = m_Q\left(1+ \frac{\tilde{\lambda}_1^2}{8}\frac{v^4}{\Lambda^2m_Q^2}\right).
\end{equation}
This linear EFT construction can be mapped to a chiral nonlinear EFT describing a composite Higgs model by identifying the cut-off $\Lambda \sim 4\pi f$ and the coupling strengths in the composite Higgs model of the order of $\tilde{y}/4\pi$, $\tilde{\lambda}/4\pi$. 
Each of the interaction terms in \eqref{EFT_1} and \eqref{EFT_2} play significant role in the VLQ phenomenology as described below.
\begin{itemize}
    \item Interaction vertices with strength $\lambda_R$ give dominant contribution to the 2-body decays of the VLQs, $T_{2/3} \to t S^0$, $X_{5/3} \to t S^+$, $Y_{8/3} \to t S^{++}$.
    \item The couplings $\Tilde{y}_t$ and $\Tilde{y}_b$ contribute to the decay widths $S^0 \to t\Bar{t}$, and $S^0 \to b\Bar{b}$, respectively, while both of them contribute to $S^+ \to t \Bar{b}$ decay width.
    \item  The mass mixing between $t$ and $T_{2/3}$ arises from the interaction term with coefficient $\Tilde{\lambda}_1$. 
    \item The coefficient $\Tilde{\lambda}_2$ contributes to the partial decay widths of $X_{5/3}\to b S^{++}$, and $T_{2/3}\to b S^{+}$, however, these decay channels are typcally sub-leading.
\end{itemize}
The benchmark parameters shown in \cref{coups_BP} can be obtained by choosing $\lambda_R\sim 1$ and $\tilde{y}_{t,b}/\Lambda\sim\tilde{\lambda}_{1,2}/\Lambda\sim 1~{\rm TeV}^{-1}$ in \cref{EFT_1,EFT_2}.

\clearpage
\section{Additional results from the recast}
\label{app:recastSR}

\begin{figure}[h!]
\begin{center}
    	\includegraphics[width=\textwidth]{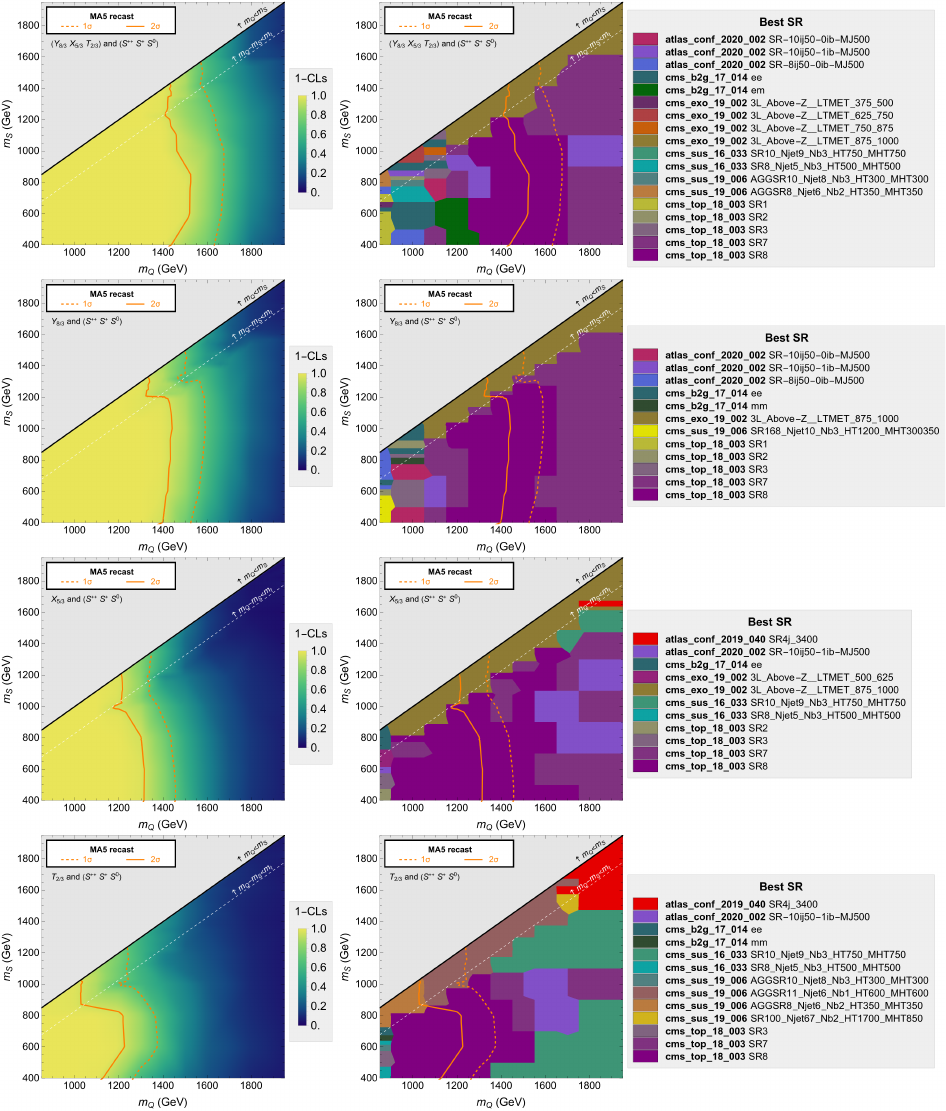}
\end{center}    	
    \caption{\sf\it Exclusion confidence levels (left column) and best SRs for each point of the scan (right column) for: combined signal $T_{2/3}$+$X_{5/3}$+$Y_{8/3}$ (top row), only $Y_{8/3}$ (second row), only $X_{5/3}$ (third row), and only $T_{2/3}$ (bottom row).}
    \label{fig:exclusion_overview}
\end{figure}

The exclusion confidence levels and the best SRs corresponding to each point of the scan for the individual VLQs and for their sum are shown in \cref{fig:exclusion_overview}.

\clearpage
\section{The non-degenerate case}
\label{app:nondegenerate}

In this appendix we show results corresponding to a scenario where VLQs and scalars are not degenerate. This is motivated by the fact that benchmarks with exact degeneracy might seem too tuned, due to a too strong assumption. We performed the analysis for the BPL benchmark where we artificially split the VLQs and scalars by 50 GeV: specifically, the central particles of the triplets, $X_{5/3}$ and $S^+$, have the nominal masses of 1700 GeV and 600 GeV respectively, while those with higher hypercharge, $Y_{8/3}$ and $S^{++}$, are lighter, with masses of 1650 GeV and 550 GeV respectively, and the particles with lower hypercharge, $T_{2/3}$ and $S^0$ are heavier, 1750 GeV and 650 GeV respectively. The decay patterns of the particles do not change significantly with respect to BPL even if masses are different. Nevertheless, we have accounted for such changes in our results. Indeed, the mass splitting has been chosen to avoid the introduction of new decays mediated by on-shell $W$'s or $Z$'s. A splitting of some tens of GeV can indeed be expected when considering loop corrections to the masses, and a rough estimation of electroweak and QCD contributions to the mass differences confirms that differences are in the ballpark of 50 GeV. We did not perform a more accurate analysis for the sake of simplifying our phenomenological estimate. 

\begin{figure}[h!]
\begin{center}
    	\includegraphics[width=\textwidth]{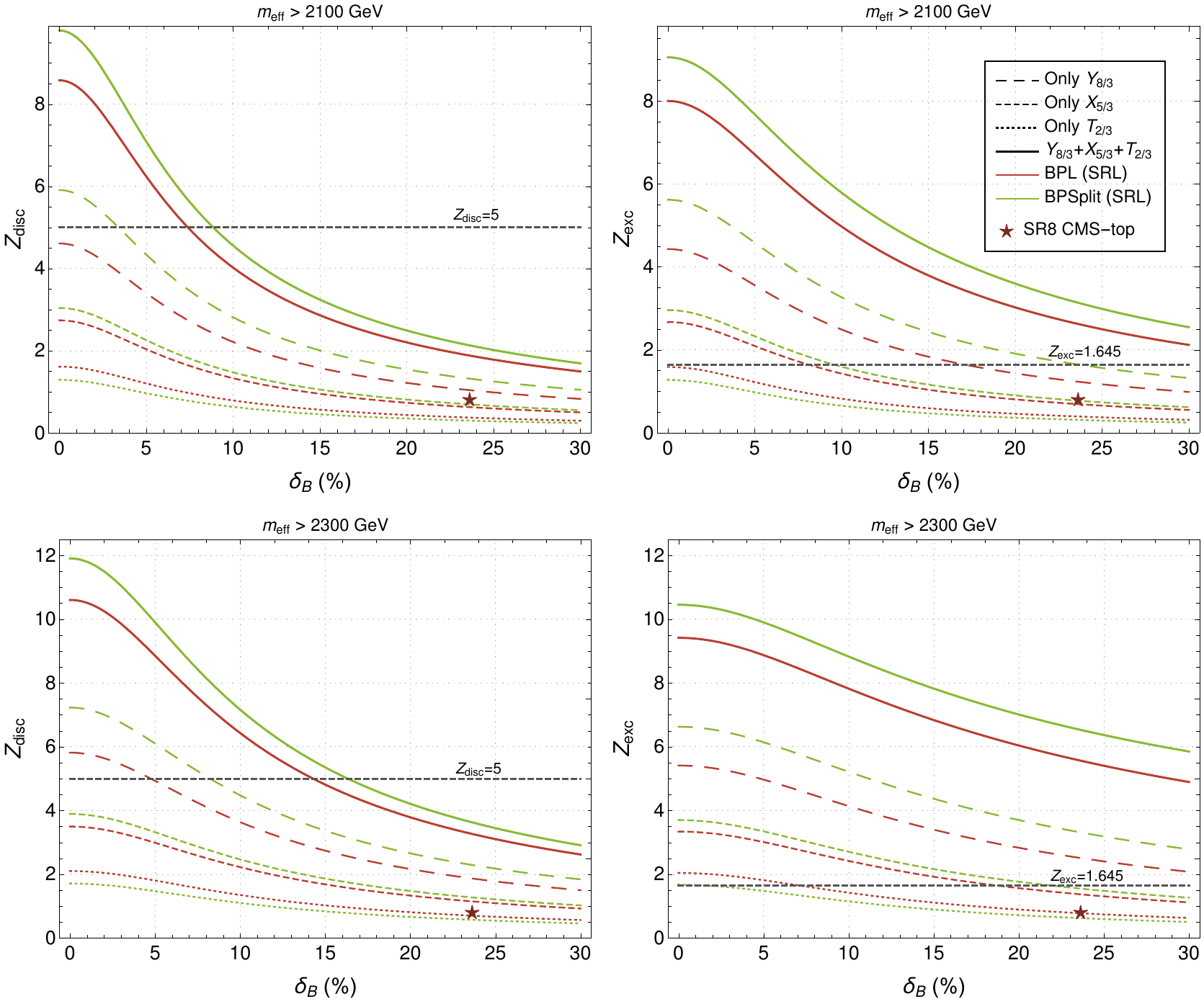}
\end{center}    	
    \caption{\sf\it Expected significance for discovery (left) and exclusion (right) at the HL-LHC are shown in SRL, for the cases with degenerate masses (red) and with a mass split (green).}
    \label{fig:projection_summary_split}
\end{figure}

Our results are shown in \cref{fig:projection_summary_split}, where it is possible to see that as one could expect, significances associated with the lightest particle, $Y_{8/3}$ increase due to the higher cross-section, and decrease for the heavier $T_{2/3}$. Overall, the combined significances increase in the whole range of systematic uncertainties, which can be expected since combined limits are mostly driven by $Y_{8/3}$. 
If the sign of mass splittings was inverted one could expect an overall worsening of limits, but results would not be significantly different compared to the degenerate case. Thus, the conclusions obtained in our simplified benchmark scenarios can indeed be used to describe the phenomenology of multiple VLQs in the same mass range.

% % % % % % % % % % % % % % % % % % % % % % % % % % % % % % % % % % % % % %
\bibliography{literature.bib}
\bibliographystyle{JHEP}
\end{document}